\documentclass[aps,pra,twocolumn,superscriptaddress,showpacs,floatfix]{revtex4}
\usepackage{graphicx}

\begin{document}

\title{Faraday waves in binary non-miscible Bose-Einstein condensates}

\author{Antun Bala\v{z}}
\email{antun@ipb.ac.rs}
\affiliation{Scientific Computing Laboratory, Institute of Physics Belgrade,
University of Belgrade, Pregrevica 118, 11080 Belgrade, Serbia}

\author{Alexandru I. Nicolin}
\email{nicolin@theory.nipne.ro}
\affiliation{\textquotedblleft{}Horia Hulubei\textquotedblright{} National Institute
for Physics and Nuclear Engineering (IFIN-HH),\\ P.O.B. MG-6, 077125, Romania}

\begin{abstract}
We show by extensive numerical simulations and analytical variational calculations that
elongated binary non-miscible Bose-Einstein condensates subject to periodic
modulations of the radial confinement exhibit a Faraday instability
similar to that seen in one-component condensates. Considering the
hyperfine states of $^{87}$Rb condensates, we show that there are
two experimentally relevant stationary state configurations: the one in which the components
form a dark-bright symbiotic pair (the ground state of the system),
and the one in which the components are segregated (first excited state).
For each of these two configurations, we show numerically that far from
resonances the Faraday waves excited in the two components are
of similar periods, emerge simultaneously, and do not impact the dynamics
of the bulk of the condensate. We derive analytically the period of
the Faraday waves using a variational treatment of the coupled Gross-Pitaevskii
equations combined with a Mathieu-type analysis for the selection
mechanism of the excited waves. Finally, we show that for a modulation
frequency close to twice that of the radial trapping, the emergent
surface waves fade out in favor of a forceful collective mode that
turns the two condensate components miscible.
\end{abstract}


\maketitle

\section{Introduction}

The excitation of surface waves through parametric resonances is one
of the oldest pattern-forming processes that goes back to Ernst Chladni's
{}``beautiful series of forms assumed by sand, fillings, or other
grains when lying upon vibrating plates'' that {}``are so striking
as to be recalled to the minds of those who have seen them by the
slightest reference'', Hans Christian \O{}rsted's experiments with
lycopodium light powders, and Michael Faraday's {}``crispations''
seen in \textquotedblleft{}fluids in contact with vibrating surfaces\textquotedblright{}
\cite{Faraday1831}. The prototypical example of parametric wave excitation
is that of a shallow disc of a liquid rigidly oscillated in the vertical
direction. In this setting the acceleration periodically modulates
the effective gravity and for drives of sufficiently large amplitudes
a surface wave instability occurs with frequency one half that of
the drive \cite{CrossHohenberg}. Such surface waves are termed Faraday waves, and the corresponding pattern-forming phenomenon
has been seen in numerous Newtonian and non-Newtonian fluids, colloidal
suspensions, ferromagnetic bodies, and, more recently, in superfluids. 

After a series of inceptive studies on extended parametric resonances
in confined superfluids \cite{PR-orig,StaliunasOne-1,StaliunasTwo-1,paramresOLorig,paramresOL,ModugnioPF},
the experimental observation of Faraday waves in $^{4}$He cells \cite{FaradayHe}
and $^{87}$Rb cigar-shaped Bose-Einstein condensates (BECs) \cite{EngelsFaraday}
catalyzed the interest in the nonlinear dynamics of parametrically-driven
ultracold gases \cite{PanosBook}. In addition to the ever-present collective oscillation modes of BECs \cite{hu1}, their parametric driving can generate soundlike density waves, which are analogous to Faraday surface waves. BEC systems also exhibit resonances, and when the driving frequency is close to one of them, forceful resonant waves can develop, eventually taking over the dynamics of the system. Furthermore, inherent nonlinearity of such systems brings about a diverse set of related dynamical phenomena and leads to their complex interplay.

Surveying the recent literature
for bosonic systems one notices the theoretical investigations into the soliton management in periodic systems \cite{MalomedBook, hadzievski},
the emergence and suppression of Faraday patterns \cite{AlexPanos,FaradayTwoComp,PhysACArina,santos,capuzzidoi},
the parametric excitation of resonances \cite{hu2} and ``scars'' in BECs \cite{katz}, spatially and temporally driven atomic interactions in optical lattices \cite{gaul1, sekh}, quantized vortices induced by spatio-temporally modulated interaction \cite{wang}, stability and decay of Bloch oscillations in the presence of time-dependent nonlinearity \cite{gaul2,gaul3}, and,
quite interestingly, the removal of excitations in BECs subject to
time-dependent periodic potentials \cite{StaliunasSuppresion}. On
a related topic, the formation of density waves has been predicted
for expanding condensates \cite{densitymodulationsSalasnich,Imambekov}
and the spontaneous formation of density waves has been recently reported
for antiferromagnetic BECs \cite{SpontPattFormBEC}. Faraday waves
have been analyzed in detail also in superfluid fermionic gases \cite{Capuzzi,Tang}
and it has been shown that the collective modes of 1D fermionic systems
can be amplified by parametric resonances to the extent of
observing a clear spin-charge separation \cite{graf}. 

Parametric resonances in BECs are usually achieved through periodic
modulation of the frequency of the trapping potential, as is the case
in Ref. \cite{EngelsFaraday}, but the recent experiments on the collective
modes of a trapped $^{7}$Li BEC through periodic modulation of the
scattering length \cite{Pollack} have opened a new direction for both experimental and theoretical \cite{Ivana}
investigations. Simultaneous modulations of the strength of the confining
potential and of the scattering length, in particular, are largely
unexplored. They can give rise to new recipes for pattern formation and, possibly, to new types of patterns.

In this paper we study the excitation of waves through periodic modulations
of the radial confinement of binary non-miscible BECs and show that
these condensates exhibit a Faraday instability similar to that seen
in one-component systems. Considering the hyperfine states of realistic $^{87}$Rb condensates which can be readily produced \cite{Jovana},
such as $\left|1, -1\right\rangle$, $\left|2, 0\right\rangle$ and $\left|1, -1\right\rangle$, $\left|2, 1\right\rangle$ pairs,
we show that there are two distinct experimentally relevant stationary configurations: one
where the components form a dark-bright symbiotic pair, which is the
ground state of the system, and one where the components are segregated, which is the first excited state of the system. Far from resonances
we show numerically for each configuration that the excited waves
are of similar periods and emerge simultaneously, and analytically find
the dispersion relation using a variational treatment in conjunction
with a Mathieu-type analysis. Finally, the resonant excitation of
collective modes is analyzed in detail.

The rest of the paper is structured as follows. In Sec.~\ref{sec:num} we describe
the numerical treatment of the coupled Gross-Pitaevskii equations
(GPEs) that describe the $T=0$ dynamics of the condensate and introduce
the two types of stationary configurations, while in Sec.~\ref{sec:var} we derive
the corresponding variational equations and the associated dispersion
relations. In Sec.~\ref{sec:res} we present our numerical and analytical results
for realistic $^{87}$Rb condensates, along with suggestions for
future experiments. The last section contains conclusions and outlook.

\section{Stationary states}
\label{sec:num}

The dynamics of binary condensates at $T=0$ is governed by the time-dependent
version of the coupled GPEs, which read \cite{Book}
\begin{eqnarray}
i\frac{\partial\psi_{j}}{\partial t} & = & -\frac{1}{2}\Delta\psi_{j}+V({\bf r},t)\psi_{j}+N_{j}U_{j}\left|\psi_{j}\right|^{2}\psi_{j}\nonumber \\
 &  & +N_{3-j}\tilde{U}\left|\psi_{3-j}\right|^{2}\psi_{j},\label{eq:GPRef}
\end{eqnarray}
where $j\in\{1,2\}$, $N_j$ is the fixed number of atoms in each component, $U_{j}=4\pi a_{j}$, $\tilde{U}=4\pi\tilde{a}$, with $a_j$ being the intra-component scattering lengths, while $\tilde{a}$ is the inter-component scattering length. Here we use numerical values similar to the experimental ones from Refs.~\cite{scattlengths, DarkBrightPhysLettA, segprl},
\begin{equation}
\label{eq:avalues}
a_{1}=100.4\, a_0\, ,\quad a_{2}=98.98\, a_0\, ,\quad \tilde{a}=a_1\, ,
\end{equation}
where $a_0$ is the Bohr radius. Small variations in scattering lengths yield similar results, but we have chosen these particular values since they correspond to a clearly non-miscible configuration at the mean-field level, thereby emphasizing the forcefulness of the miscibility transition, which represents one of our main motivations, as we will see in Sec.~\ref{sec:res}. For simplicity, we use natural units $\hbar=m=1$ throughout the paper, and component normalization
\begin{equation}
\int d{\bf r}\left|\psi_{j}({\bf r},t)\right|^{2}=1\, .\label{eq:Normalization}
\end{equation}

\begin{figure}[!b]
\begin{center}
\includegraphics[width=8.5cm]{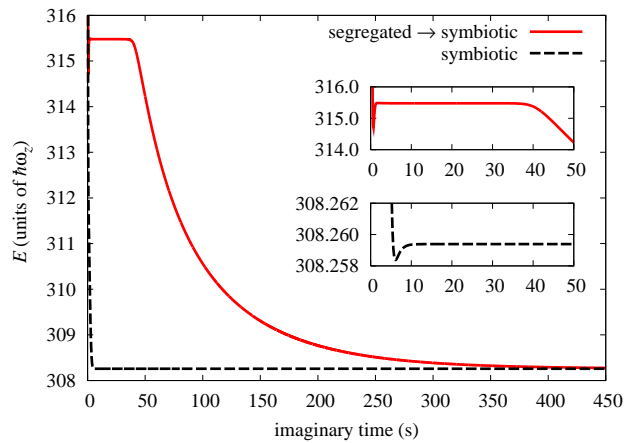}
\caption{(Color online) Typical imaginary-time propagation of the total energy for a two-component condensate system
with $N_{1}=2.5\cdot 10^5$ atoms in the state A and $N_{2}=1.25\cdot 10^5$ in the state B, $\Omega_\rho=160\times 2\pi$~Hz, $\Omega_z=7\times 2\pi$~Hz. The dashed black curve corresponds to the evolution from initial wave functions set to two identical Gaussians, which immediately yields the ground state, while the full red curve corresponds to evolution from the initial wave functions set to two well-separated Gaussians, yielding the first excited state of the system, which eventually decays to the ground state.}
\label{fig1}
\end{center}
\end{figure}

The stationary states of binary Bose-Einstein condensates are obtained by solving the time-independent version of Eqs.~(\ref{eq:GPRef}), in which terms with time derivatives of wave functions are replaced by the terms containing
chemical potentials $\mu_j$ of components,
\begin{eqnarray}
-\frac{1}{2}\Delta\psi_{j}+V({\bf r},t)\psi_{j}+N_{j}U_{j}\left|\psi_{j}\right|^{2}\psi_{j}\nonumber \\
+N_{3-j}\tilde{U}\left|\psi_{3-j}\right|^{2}\psi_{j} & = & \mu_{j}\psi_{j}.\label{eq:GPTimeInDep}
\end{eqnarray}
In this paper we consider the two hyperfine states of $^{87}$Rb (hereafter referred to as states A and B) in an external potential of the form
\begin{equation}
V({\bf r},t)=\frac{1}{2}\Omega_\rho^{2}(t)\rho^{2}+\frac{1}{2}\Omega_{z}^{2}z^{2},\label{eq:ExtPot}
\end{equation}
where $\rho^{2}=x^{2}+y^{2}$, such that the system is cylindrically-symmetric, i.e. $\psi_j({\bf r},t)\equiv \psi_j(\rho, z, t)$, and GPEs are
effectively two-dimensional. We also assume that the system is highly elongated and strongly confined in the radial direction, $\Omega_\rho(t)\gg\Omega_{z}$.
The above time-dependent system of GPEs (\ref{eq:GPRef}), as well as its time-independent counterpart Eq.~(\ref{eq:GPTimeInDep}), can be solved numerically using various approaches \cite{pla-manybody, balazpre, diag1, diag2, becpla}. Here we choose the efficient split-step Crank-Nicolson approach developed by Adhikari and Muruganandam in Ref.~\cite{AdhikariCode}. We have implemented this algorithm in the C programming language, and use it for all numerical simulations presented here.

Binary condensates exhibit a wide range of interesting and experimentally relevant configurations,
which go from the one-dimensional soliton pairs tabulated in Ref.~\cite{2CompSolPairs},
to the more exotic two-dimensional vortex-lattices \cite{2CompVorLatt},
and vortex-bright-soliton structures \cite{2CompVortexBrightSol}.
From the mean-field theory it follows that binary condensates are miscible if the condition
$\tilde{U}<\sqrt{U_{1}U_{2}}$ is satisfied \cite{AoChui}. For the values from Eq.~(\ref{eq:avalues}) this condition is not satisfied, and the system is non-miscible \cite{scattlengths}.

The stationary states are computed numerically by imaginary-time propagation
until we achieve the convergence of wave functions and physical quantities of the system,
chemical potentials and mean-square radii of components, and the total system energy. Fig.~\ref{fig1} illustrates the results obtained using this approach for different initial conditions. It shows total energy of the system as a function of the imaginary time. Using the various initial conditions, we were able to numerically calculate two relevant stationary configurations: the one in which the components form a
dark-bright symbiotic pair (ground state), and the one in which the components are spatially well segregated. From Fig.~\ref{fig1} we can see that the segregated state is a first excited state of the system, as it decays to the ground state after sufficiently long imaginary-time propagation.

\begin{figure}[!t]
\begin{center}
\includegraphics[width=7.8cm]{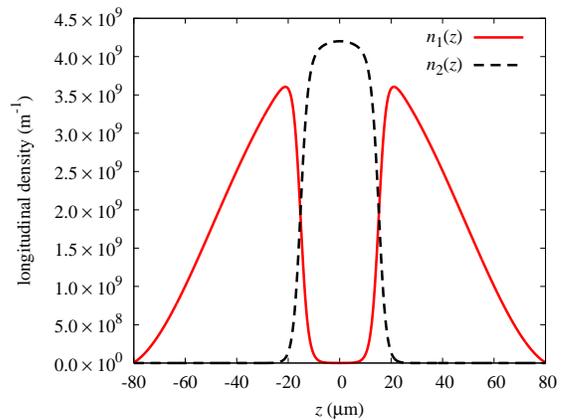}
\caption{(Color online) Typical longitudinal density profiles for a condensate
in the ground state (dark-bright symbiotic pair). The parameters are the same as in Fig.~\ref{fig1},
the full red curve shows the density profile for atoms in the state A, the dashed black line for atoms in the state B.}
\label{fig2}
\end{center}
\end{figure}

The first type of stationary solutions supported by the GPEs, the ground state, consists
of a symmetric, Thomas-Fermi-type density profile with a hole
in the center in one component, and a well-localized density
peak positioned in the center of the trap in the other component.
Fig.~\ref{fig2} shows typical longitudinal density profiles
\begin{equation}
n_{j}(z)=\int_{0}^{\infty}d\rho\ 2\pi \rho \left|\psi_{j}(\rho,z)\right|^{2}\label{eq:LongDensProf}
\end{equation}
of the two components for the ground state, obtained through the imaginary-time propagation,
starting from the two identical Gaussian initial states.

We note the apparent similarity of the obtained dark-bright symbiotic solutions and the dark-bright soliton molecules, known to
exist in homogenous systems. These soliton molecules were originally
predicted in a nonlinear optics setting \cite{DBThOne,DBThTwo} and
were first observed in photorefractive crystals \cite{DBSolitonsOptics}.
Following their exposure to the BEC community \cite{PredictionDBOrigTheor},
they were observed experimentally using the $^{87}$Rb condensates \cite{DBSolitonFirstExp},
and were subsequently addressed theoretically in Refs.~\cite{NistazakisDB,DarkBrightPhysLettA,PerezGarciaDBSoliton}.
In homogeneous condensates the dark component effectively acts as a
trapping potential for the bright component through the nonlinear interaction,
and this mechanism is well preserved in inhomogeneous systems, albeit
the dark component is cut-off away from the center of the trap and
the bright component tends to be more narrow due to the additional
confinement by the trap. However, symbiotic solutions from Fig.~\ref{fig2} do not represent inhomogeneous counterparts of soliton molecules, since they are purely real, and the first component does not exhibit a jump of $\pi$ in the phase between its left and right part.

\begin{figure}[!t]
\begin{center}
\includegraphics[width=7.8cm]{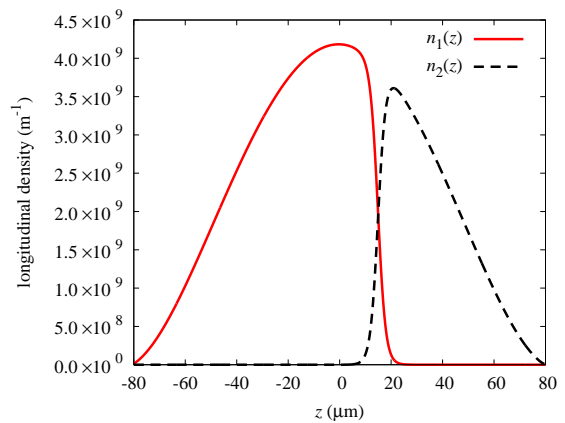}
\caption{(Color online) Typical longitudinal density profiles for a condensate
in the first excited state (segregated state). The parameters are the same as in Fig.~\ref{fig1},
the full red curve shows the density profile for atoms in the state A, the dashed black line for atoms in the state B.}
\label{fig3}
\end{center}
\end{figure}

The second type of stationary solutions supported by the GPEs, the first excited state, is presented in
Fig.~\ref{fig3}. It consists of two opposing, asymmetric Thomas-Fermi-type density profiles,
with the negligible overlap. This segregation of the components stems from the repulsive character
of the interaction and the non-miscible nature of the considered binary condensate. Naturally, there exist a large number of further excited states of the system that can be calculated numerically as well, but we focus only on the two relevant states, shown in Figs.~\ref{fig2} and \ref{fig3}.

Experimentally, such states can readily be realized using current technology. While our analysis considers elongated traps approaching a one-dimensional limit, we note that the state in Fig.~\ref{fig2} has already been observed in three-dimensional traps where a magnetically trapped BEC of $^{87}$Rb atoms was prepared in a mixture of the $\left|1,-1\right\rangle$ and $\left|2,1\right\rangle$ hyperfine states and was observed to assume a ball-shell structure \cite{cornell, hall}. By introducing differential trap shifts for the two components as in Ref.~\cite{cornell}, the segregated state in Fig.~\ref{fig3} can readily be generated as well. Alternatively, optical traps can be employed to confine atoms in a state independent way, and an additional magnetic gradient can be used to separate components of a mixture. Using a miscible mixture, this phase separation in an optical trap has been generated e.g.~in Ref.~\cite{segprl} and can also be extended to optical lattice systems \cite{ketterle}. Using a microwave sweep in a high-bias magnetic field, the miscible mixture in the latter experiments can be converted into an immiscible mixture by changing the hyperfine state of one of the components.

At the end, let us also note that the stationary character of the two numerically calculated states presented here was cross-checked by performing a real-time propagation for long time intervals (much longer than those considered in the rest of the paper). While in the imaginary-time propagation the first excited state eventually decays to the ground state, when real-time propagation is performed both stationary states are found to be stable for all practical purposes.

\section{Variational treatment}
\label{sec:var}

In this section we develop variational analytic approach suitable for study of the emergence and characterization of Faraday patterns in non-miscible binary condensates, induced by harmonic modulation of the radial trapping frequency. For each of the two stationary states calculated in Sec.~\ref{sec:num}, we propose a suitable variational ansatz for component wave functions, approximately solve the ensuing equations, and analytically derive expressions for periods of Faraday waves induced far from resonances. The obtained analytical results are compared with the numerical results in Sec.~\ref{sec:res}.

\subsection{Symbiotic pair state}
\label{sec:var-ss}

To construct a suitable variational ansatz for component wave functions in the case of the symbiotic pair state, we consider an equivalent one-component scenario, in which
the bright state $\psi_2$ evolves in the combined effective field created by
the dark state $\psi_1$ and by the trapping potential. To further simplify the
problem, in this subsection we will also assume that the values of three scattering lengths from Eq.~(\ref{eq:avalues}) are equal,
namely $a_{1}=a_{2}=\tilde{a}\equiv a=100.4\, a_0$, or $U_1=U_2=\tilde U\equiv U$, which is justified since the values measured experimentally are quite close \cite{scattlengths}.

In this setting, the effective one-component Lagrangian density for the bright state $\psi_2$ is given by 
\begin{eqnarray}
{\cal L}_2(\rho,z,t) & = & \frac{i}{2}\left(\psi_{2}\frac{\partial\psi_{2}^{*}}{\partial t}-\psi_{2}^{*}\frac{\partial\psi_{2}}{\partial t}\right)+\frac{1}{2}\left|\nabla\psi_{2}\right|^{2}\nonumber \\
 &  & +V(\rho,z,t)\left|\psi_{2}\right|^{2}+\frac{UN_{2}}{2}\left|\psi_{2}\right|^{4}\nonumber \\
 &  & +UN_{1}\left|\psi_{1}\right|^{2}\left|\psi_{2}\right|^{2},\label{eq:LagrangianTypeB}
\end{eqnarray}
where we have chosen the following ans\"atze for component wave functions:
\begin{eqnarray}
\psi_{1}(\rho,z,t) & = & {\cal N}_1\exp\left(-\frac{\rho^{2}}{2w_\rho^{2}(t)}+i\rho^{2}\alpha^{2}(t)\right)\nonumber \\
 &  & \times\left[1-\exp\left(-\frac{z^{2}}{2w_{z}^{2}}\right)\right]\, ,\label{eq:Psi1TypeB}\\
\psi_{2}(\rho,z,t) & = & {\cal N}_2\exp\left(-\frac{\rho^{2}}{2w_\rho^{2}(t)}-\frac{z^{2}}{2w_{z}^{2}}+i\rho^{2}\alpha^{2}(t)\right)\nonumber \\
 &  & \times\big[1+(u(t)+iv(t))\cos kz\big]\, ,\label{eq:Psi2TypeB}
\end{eqnarray}
with straightforward interpretation for the variational parameters: the radial and the longitudinal bright state widths $w_\rho(t)$ and $w_z$, the phase $\alpha^2(t)$, and the complex amplitude $u(t)+iv(t)$ of the Faraday wave in the bright component, with the period $2\pi/k$. Note that, in order to keep the analytics reasonably simple and tractable, we assume that the longitudinal condensate width $w_z$ is constant, and include the surface wave only in the bright
component, in a manner similar as in Ref. \cite{ProcRomAcad,AlexRRP2011,AlexPhysA2012}. As the
role of the dark component is mainly that of an additional trapping potential,
these simplification of ans\"atze have little impact on the final results.
The normalization factors ${\cal N}_j$ are calculated from normalization conditions 
\begin{eqnarray}
\int_{-L}^L dz\ \int_{0}^{\infty}d\rho\ 2\pi \rho\left|\psi_{1}(\rho,z,t)\right|^{2}&=&1\, ,\label{eq:NormPsi1TypeB}\\
\int_{-\infty}^\infty dz\ \int_{0}^{\infty}d\rho\ 2\pi \rho\left|\psi_{2}(\rho,z,t)\right|^{2}&=&1\, ,\label{eq:NormPsi2TypeB}
\end{eqnarray}
where $2L$ is the longitudinal spatial extent of the dark state component, obtained
by solving the stationary system of GPEs and assumed to be constant.

After inserting the ans\"atze (\ref{eq:Psi1TypeB}) and (\ref{eq:Psi2TypeB}) into the Lagrangian density (\ref{eq:LagrangianTypeB}), we calculate the Lagrangian and derive the following variational equations for the parameter functions $w_\rho(t)$, $\alpha(t)$, $u(t)$ and $v(t)$:
\begin{eqnarray}
\dot{w}_\rho & = & 2w_\rho\alpha,\label{eq:wrTypeB}\\
\dot{\alpha} & = & \frac{1}{2w_\rho^{4}}-\frac{\Omega_\rho^{2}}{2}-2\alpha^{2}\nonumber \\
 &  & +\frac{U\left(3\sqrt{8}LN_{2}+g_\alpha\sqrt{\pi}w_z\right)}{24\pi^{\frac{3}{2}}w_\rho^4w_z\left[2L-\left(\sqrt{8}-1\right)\sqrt{\pi}w_{z}\right]},\label{eq:alphaTypeB}\\
\dot{u} & = & \frac{k^{2}v}{2},\label{eq:uTypeB}\\
\dot{v} & = & -\frac{k^{2}u}{2}-\frac{UN_2u}{\sqrt{2}\pi^\frac{3}{2}w_{z}w_\rho^{2}},\label{eq:vTypeB}
\end{eqnarray}
where $g_\alpha=(12+6\sqrt{2}-8\sqrt{6})N_1-3(4-\sqrt{2})N_2.$ In
addition to these ordinary differential equations, we have an algebraic
equation for $w_{z}$, namely 
\begin{eqnarray}
\label{eq:wzTypeBAlgebraic}
0&=&1+\left(\frac{9}{4}-\sqrt{2}\right)\frac{\pi w_{z}^{2}}{L^{2}} -\frac{(2\sqrt{2}-1)\sqrt{\pi}w_{z}}{L}\\
 &&\hspace*{-5mm}-\frac{Uw_{z}N_2\pi^{-\frac{3}{2}}}{w_{\rho 0}^{2}\left(\Omega_{z}^{^{2}}w_{z}^{2}-1\right)}\left[2^\frac{5}{2}-\frac{(16-2^\frac{5}{2})\sqrt{\pi}w_{z}}{L}-\frac{\pi g_z w_{z}^{2}}{3N_2L^{2}}\right],\nonumber
\end{eqnarray}
where $g_z=2(9\sqrt{2}-16\sqrt{3}+4\sqrt{6}+6)N_1-3(9\sqrt{2}-8)N_2$ and  $w_{\rho 0}=w_\rho(0)$.
The tacit assumption that the condensate is of constant longitudinal
extent $2L$ will be further justified in the next section, where we present the
full numerical results.

Let us first emphasize that, as will be shown later numerically, the dynamics of the bulk of the condensate, in the first approximation,
is not impacted by that of the surface wave.
Eqs.~(\ref{eq:wrTypeB}) and (\ref{eq:alphaTypeB}) are, in fact,
similar to those derived in Ref. \cite{CalcVar} for the collective
dynamics of low-density one-component condensates, while Eqs.~(\ref{eq:uTypeB}) and (\ref{eq:vTypeB}) resemble those derived in
Refs. \cite{AlexPanos,PhysACArina} for Faraday waves in low- and
high-density one-component condensates.
Second, we stress that
for a modulated radial trapping $\Omega_\rho(t)=\Omega_{\rho 0}\cdot(1+\epsilon\sin\omega t)$, Eqs.~(\ref{eq:wrTypeB}) and (\ref{eq:alphaTypeB}) exhibit a series of
parametric resonances for $\omega=\Omega_{\rho 0}$ (self-resonance) and $\omega=2\Omega_{\rho 0}/n^{2}$, where $n$ is an integer  \cite{PR-orig}. The widest resonance is that at $\omega=2\Omega_{\rho 0}$ (for $n=1$), followed by the self-resonance at $\omega=\Omega_{\rho 0}$, which has been evidenced in Ref.~\cite{EngelsFaraday} for one-component $^{87}$Rb condensates. The other ($n\geq 2$) resonances are very narrow and are of little experimental interest, since for such low frequencies the excited
waves have periods comparable to the longitudinal extent of the condensate and are therefore difficult to observe.

Finally, let us note that, unlike in one-component condensates, where Faraday waves emerge rapidly enough as to hide the resonant behavior at $\omega=2\Omega_{\rho 0}$ \cite{EngelsFaraday}, in two-component systems the forceful resonant behavior is dominant, as we will show numerically in Sec.~\ref{sec:res}. 

Far from resonances we can approximately calculate the radial width as
\begin{equation}
w_\rho\approx\Omega^{-\frac{1}{2}}_\rho\left(1+\frac{U(3\sqrt{8}LN_2+g_\alpha\sqrt{\pi}w_z)}{12\pi^{\frac{3}{2}}w_z\left[2L-\left(\sqrt{8}-1\right)\sqrt{\pi}w_{z}\right]}\right)^\frac{1}{4},\label{eq:wrsolappTypeB}
\end{equation}
which stems from Eqs.~(\ref{eq:wrTypeB}) and (\ref{eq:alphaTypeB})
by neglecting $\ddot{w}_\rho(t)$, and casts the equation for $u(t)$ into the form
\begin{equation}
\ddot{u}(\tau)+u(\tau)\left[a(k,\omega)+\epsilon b\left(k,\omega\right)\sin2\tau\right]=0\, ,\label{eq:MathieuTypeB}
\end{equation}
A dimensionless time $\tau$ is introduced as $\omega t=2\tau$, and
\begin{eqnarray}
a(k,\omega) & = & \frac{k^{4}}{\omega^{2}}+ \frac{k^{2}}{\omega^{2}}\, \Lambda_\mathrm{sym}\, ,\label{eq:akwTypeB}\\
b(k,\omega) & = & \frac{k^{2}}{\omega^{2}}\, \Lambda_\mathrm{sym}\, ,\label{eq:bkwTypeB}\\
\hspace*{-2mm}
\Lambda_\mathrm{sym}&=&  \frac{3\sqrt{8g_{ab}}UN_{2}\Omega_{\rho 0}}{\sqrt{3\sqrt{8}ULN_{2}w_z+\sqrt{\pi}w_{z}^2\left(g_\alpha U+12\pi g_{ab}\right)}} ,\ \label{eq:LTypeB}\hspace*{4mm}\\
g_{ab}&=&2L-(2\sqrt{2}-1)\sqrt{\pi}w_{z}\, .
\end{eqnarray}
For small positive values of the radial modulation amplitude $\epsilon$ and positive values of the function $b(k,\omega)$,
Eq.~(\ref{eq:MathieuTypeB}) has solutions of the form $\exp(\pm i\mu\tau)\sin\sqrt{a}\tau$
and $\exp(\pm i\mu\tau)\cos\sqrt{a}\tau$, where $\mbox{Im}[\mu]$
consists of a series of lobes positioned around the solution of the equation $a(k,\omega)=n^{2}$,
with $n$ being an integer \cite{carteMathieu}. The lobe centered around
$a(k,\omega)=1$ is the largest, and it yields the most unstable solutions \cite{AlexPanos,PhysACArina}, determined by
\begin{equation}
\label{eq:kFsym}
k_\mathrm{F,sym} =\sqrt{-\frac{\Lambda_\mathrm{sym}}{2}+\sqrt{\frac{\Lambda_\mathrm{sym}^2}{4}+\omega^2}}\, .
\end{equation}
As these solutions have a frequency of $\omega/2$, half that of the parametric drive,
they are usually referred to as Faraday waves in honor of Faraday's
classic study \cite{Faraday1831}.

Close to a resonance \cite{song}, the approximation used above for $w_\rho(t)$ breaks down,
and one cannot generally construct the explicit equation for $u(t)$.
We know, however, that the instabilities appear due to the resonant
energy transfer between the radial mode and the surface wave, which
entails that $w_\rho(t)$ and $u(t)$ are of equal frequency. For small
values of the modulation amplitude $\epsilon$ this requires that the condition $a(k,\omega)=2^{2}$ is satisfied.

\subsection{Segregated state}
\label{sec:var-seg}

To consider the case of a segregated excited state, we build the variational equations starting from the habitual Gross-Pitaevskii
Lagrangian density \cite{Book,CalcVar}, written for a two-component condensate in the form:
\begin{eqnarray}
{\cal L}(\rho,z,t) & = & \sum_{j=1,2}\left[\frac{i}{2}\left(\psi_{j}\frac{\partial\psi_{j}^{*}}{\partial t}-\psi_{j}^{*}\frac{\partial\psi_{j}}{\partial t}\right)+\frac{1}{2}\left|\nabla\psi_{j}\right|^{2}\right.\nonumber \\
 &  & \left.+V(\rho,z,t)\left|\psi_{j}\right|^{2}+\frac{U_{j}N_{j}}{2}\left|\psi_{j}\right|^{4}\right]N_{j}\nonumber \\
 &  & +\tilde{U}N_{1}N_{2}\left|\psi_{1}\right|^{2}\left|\psi_{1}\right|^{2}\, .\label{eq:LagrangianTypeA}
\end{eqnarray}
To variationally describe the wave functions of two BEC components, we use cylindrically-symmetric ans\"atze tailored around the
usual Gaussian envelopes \cite{CalcVar} that describe the bulk of the condensate, to which we graft a surface wave \cite{ProcRomAcad,AlexRRP2011,AlexPhysA2012},
\begin{eqnarray}
\psi_{j}(\rho,z,t) & = & {\cal N}_j \exp\left(-\frac{\rho^{2}}{2w_\rho^{2}(t)}-\frac{z^{2}}{2w_{z}^{2}}+i\rho^{2}\alpha^{2}(t)\right)\nonumber \\
 &  &\hspace*{-9mm} \times\left[1+(u(t)+iv(t))\cos kz\right]\, \theta\left((-1)^j z\right)\, ,\label{AnsatzTypeA}
\end{eqnarray}
where $\theta$ represents Heaviside step function, and normalization factors are determined from
\begin{equation}
\int_{-\infty}^{\infty}dz\ \int_{0}^{\infty}d\rho\ 2\pi \rho\left|\psi_{j}(\rho,z,t)\right|^{2}=1\, .\label{eq:NormPsijTypeA}
\end{equation}
As far as the longitudinal components are concerned, the trial wave
functions consist of two opposing half-Gaussians of equal widths and
amplitudes, positioned in the center of the trap. As there
is no overlap between the two components and the period of the excited
wave is smaller than the longitudinal extent of the condensate, for small-amplitude waves the Euler-Lagrange equations can
be written as
\begin{eqnarray}
\hspace*{-6mm}
\dot{w}_\rho & = & 2w_\rho\alpha\, ,\label{wrTypeA}\\
\dot{\alpha} & = & \frac{1}{2w_\rho^{4}}-\frac{\Omega_\rho^{2}}{2}-2\alpha^{2}
+\frac{g}{\sqrt{8}\pi^{3/2}Nw_\rho^{4}w_{z}}\, ,\label{eq:alphaTypeA}\\
\dot{u} & = & \frac{k^{2}v}{2},\label{uTypeA}\\
\dot{v} & = & -\frac{k^{2}u}{2}-\frac{gu}{\sqrt{8}\pi^{3/2}Nw_\rho^{2}w_{z}}\, ,\label{vTypeA}
\end{eqnarray}
where $g=N_{1}^{2}U_{1}+N_{2}^{2}U_{2}$ and $N=N_{1}+N_{2}$.
As in the previous subsection, in writing the equations above we have assumed
that the condensate has a frozen longitudinal dynamics (apart from the dynamics of the grafted wave), and that the corresponding
width $w_z$ is constant. Its value is determined from the algebraic
equation
\begin{equation}
\frac{1}{2w_{z}^{4}}-\frac{\Omega_{z}^{2}}{2}+\frac{g}{\sqrt{8}\pi^{3/2}w_{\rho 0}^{2}w_{z}^{3}}=0\,  ,\label{eq:wzTypeA}
\end{equation}
where $w_{\rho 0}=w_\rho(0)$. As in the case of symbiotic states, we will consider modulation of the radial trapping frequency of the form $\Omega_\rho(t)=\Omega_{\rho 0}\cdot(1+\epsilon\sin\omega t)$. Following a similar reasoning, we find that the system again exhibits a series of parametric resonances: a self-resonance at $\omega=\Omega_{\rho 0}$, and series of resonances for $\omega=2\Omega_{\rho 0}/n^{2}$, where $n$ is an integer. As before, the strongest resonance is that at $\omega=2\Omega_{\rho 0}$.

Far from resonances, we can approximate the radial width as
\begin{equation}
w_\rho\approx\Omega_\rho^{-\frac{1}{2}}\left[1+\frac{g}{\sqrt{2}\pi^{3/2}Nw_{z}}\right]^{1/4}\, ,\label{eq:wrwolappTypeA}
\end{equation}
while Eqs.~(\ref{uTypeA}) and (\ref{vTypeA}) can then be conveniently
recast in the form of a Mathieu equation, 
\begin{equation}
\ddot{u}(\tau)+u(\tau)\left[a(k,\omega)+\epsilon b\left(k,\omega\right)\sin2\tau\right]=0\, ,\label{eq:MathieuTypeA}
\end{equation}
where the time $\tau$ is introduced as $\omega t=2\tau$,
\begin{eqnarray}
a(k,\omega) & = & \frac{k^{4}}{\omega^{2}} + \frac{k^2}{\omega^2}\Lambda_\mathrm{seg}\, ,\label{eq:aMathieuTypeA}\\
b(k,\omega) & = & \frac{k^2}{\omega^2}\Lambda_\mathrm{seg}\, ,\label{eq:bMathieuTypeA}
\end{eqnarray}
and
\begin{equation}
\Lambda_\mathrm{seg}  =  \frac{4g\Omega_{\rho 0}}{\sqrt{ \sqrt{2}\pi^\frac{3}{2}Nw_z g+2\pi^3N^2w_{z}^2}}\, ,\label{eq:LTypeA}
\end{equation}
As before, the most unstable solutions correspond
to $a(k,\omega)=1$, and are determined by
\begin{equation}
\label{eq:kFseg}
k_\mathrm{F,seg} =\sqrt{-\frac{\Lambda_\mathrm{seg}}{2}+\sqrt{\frac{\Lambda_\mathrm{seg}^2}{4}+\omega^2}}\, .
\end{equation}
Close to resonances, following a similar reasoning as in the case of symbiotic pair solution, we conclude that the radial mode $w_{r}(t)$ and the surface wave $u(t)$ are of equal frequency, which, for small values of the modulation amplitude $\epsilon$, is determined by solving $a(k,\omega)=2^{2}$.

\section{Results and discussion}
\label{sec:res}

To investigate the emergence of Faraday waves, we solve numerically
the coupled set of time-dependent GPEs (\ref{eq:GPRef}) for the exact values of scattering lengths from Eq.(\ref{eq:avalues}) and study the dynamics of the longitudinal density profiles of BEC components,
\begin{equation}
n_{j}(z,t)=\int_{0}^{\infty}d\rho\ 2\pi \rho \left|\psi_{j}(\rho,z,t)\right|^{2}\, , \label{eq:LongDensProfTimeDep}
\end{equation}
as well as their Fourier spectra. Our main result is that, far from resonances,
Faraday waves of similar periods appear simultaneously in both BEC components for both considered initial configurations (symbiotic pair state, segregated excited state), and that these waves have almost no effect on the dynamics of the bulk of the condensate (surface waves represent only a small perturbation of the stationary state). For
the self-resonance at $\omega=\Omega_{\rho 0}$, we show that surface waves appear considerably faster than the Faraday ones, and that, similar to the one-component case reported in Ref.~\cite{EngelsFaraday}, such resonant waves have smaller period than the Faraday waves. To understand this latter feature, we recall from the previous section
that the instability now sets in due to the resonant energy transfer between the radial mode and the surface wave, which changes the frequency of the surface wave and consequently the observed period. Finally, we numerically study the forceful resonant dynamics of the system at $\omega=2\Omega_{\rho 0}$, which has not been seen previously in one-component condensates \cite{EngelsFaraday}. This strong resonant behavior is found to facilitate the dynamical transition of the system from non-miscible to the miscible state.

\subsection{Symbiotic pair states}
\label{sec:res-ss}
\begin{figure}[!t]
\begin{center}
\includegraphics[width=6.8cm]{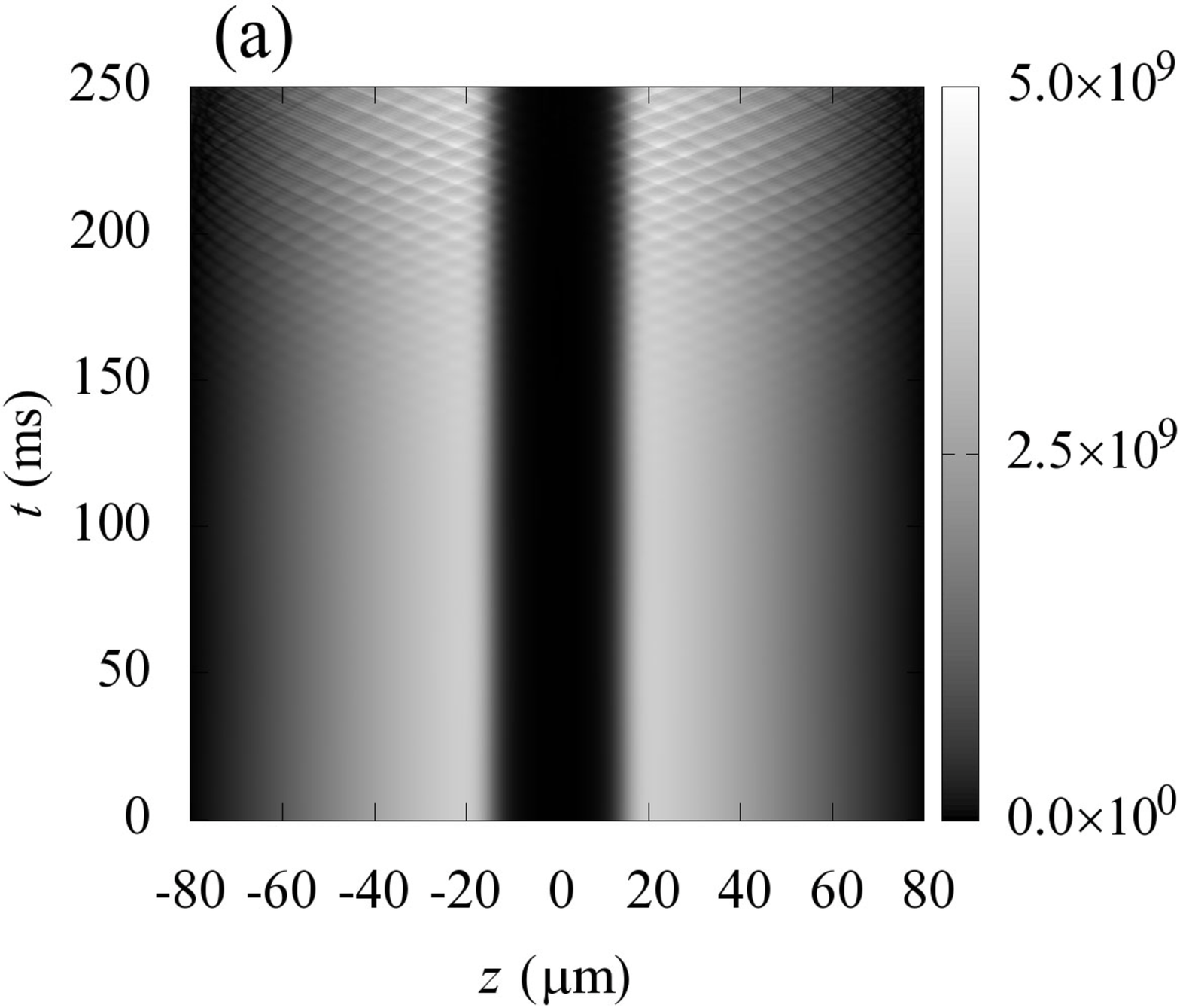}
\includegraphics[width=6.8cm]{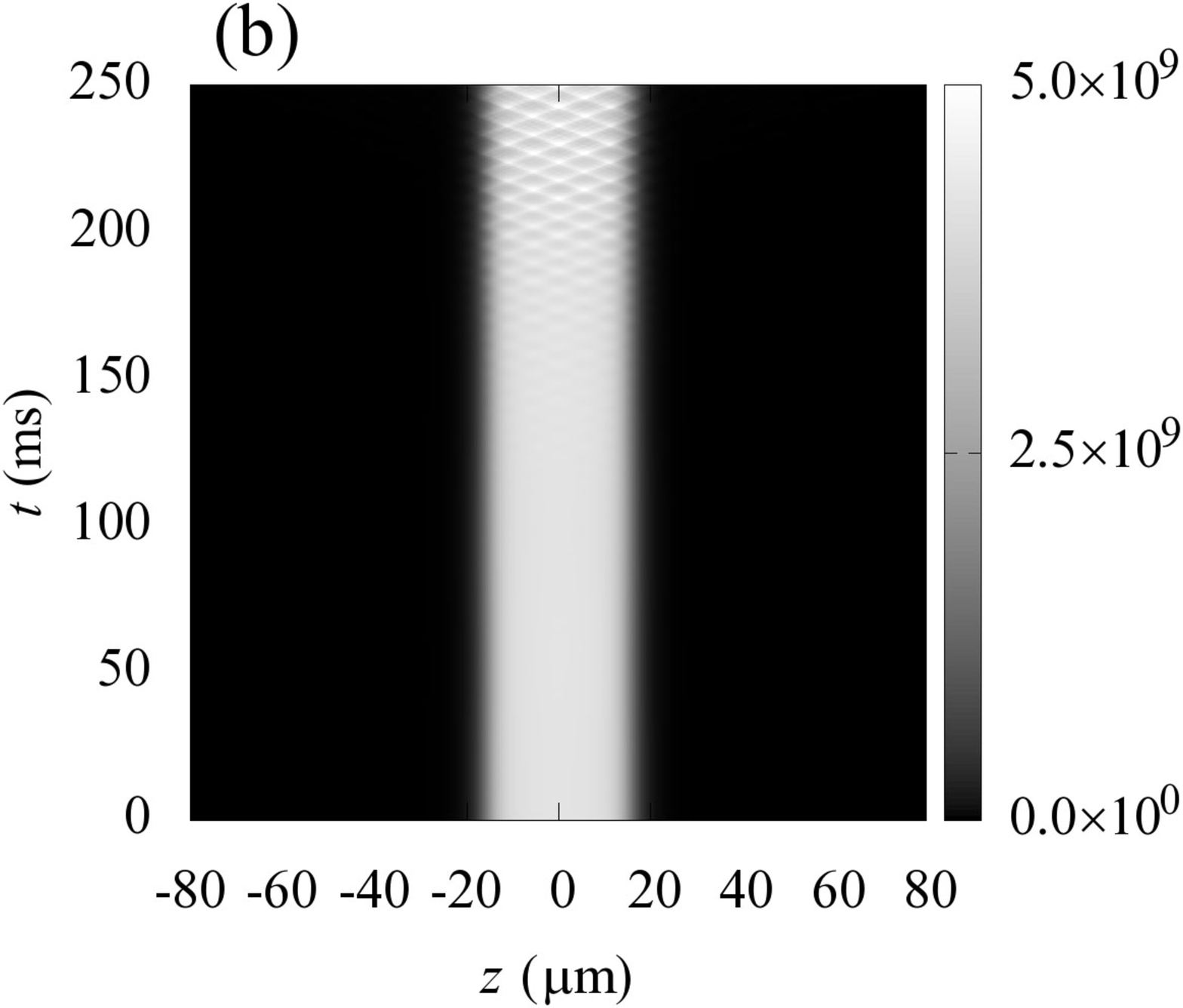}
\caption{Emergence of Faraday waves in a two-component BEC system in the real-time evolution of longitudinal density profiles: (a) $n_1(z, t)$ and (b) $n_2(z, t)$. The system is initially in the symbiotic pair state, and is modulated with the amplitude $\epsilon=0.1$ and frequency $\omega=250\times 2\pi$~Hz. The system contains $N_{1}=2.5\cdot 10^5$ atoms in the state A and $N_{2}=1.25\cdot 10^5$ atoms in the state B, confined by the trap with $\Omega_{\rho 0}=160\times 2\pi$~Hz and  $\Omega_z=7\times 2\pi$~Hz.}
\label{fig4}
\end{center}
\end{figure}
\begin{figure}[!ht]
\begin{center}
\includegraphics[width=7cm]{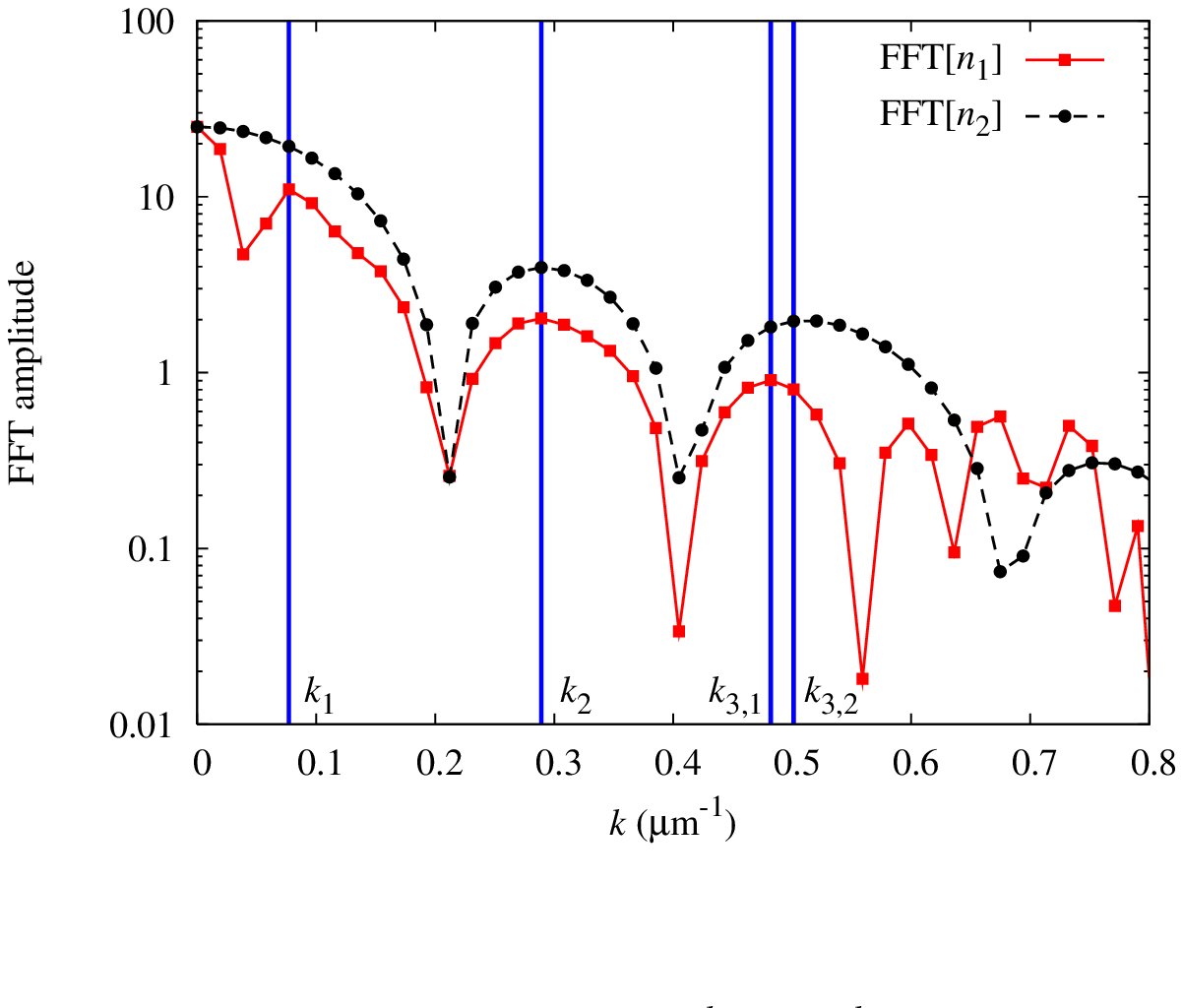}
\caption{(Color online) Fast Fourier transform of density profiles from Fig.~\ref{fig4} for two condensate components at $t=200$~ms.}
\label{fig5}
\end{center}
\end{figure}

In Fig.~\ref{fig4} we show the emergence of Faraday waves in real-time dynamics of a two-component condensate with
$N_{1}=2.5\cdot 10^5$ atoms in the state A and $N_{2}=1.25\cdot 10^5$ atoms in the state B, starting
from a symbiotic pair ground state configuration. The magnetic trap has the parameters
$\{\Omega_{\rho 0},\Omega_{z}\}=\{160\times 2\pi\, \mbox{ Hz},7\times 2\pi\, \mbox{ Hz}\}$,
and we consider the modulation frequency $\omega=250\times 2\pi$~Hz to be far from resonances. The
Faraday waves are visible after 150~ms, and we show in Fig.~\ref{fig5} the Fourier spectrum
of density profiles of the condensate at $t=200$ ms. The spectrum
exhibits several peaks, and the first two peaks, $k_1$ and $k_2$, common for both components, are related to the geometry of the system. The first peak corresponds to the period $\ell_1=2\pi/k_1=81.5\, \mu$m, the longitudinal extent of the system, while the second one corresponds to the period $\ell_2=2\pi/k_2=21.7\, \mu$m, the extent of the central dip in the fist component (or, equivalently, the extent of the second component). The periods of Faraday waves are determined by the peaks $k_{3,j}$, and have very close values, $\ell_{3,1}= 13.0\, \mu$m and $\ell_{3,2}= 12.5\, \mu$m.
The dispersion relation (\ref{eq:kFsym}) derived in Sec.~\ref{sec:var-ss} indicates a period
of $12.0, \mu$m, which is in excellent agreement with the numerical results.

This demonstrates that the variational ans\"atze from Sec.~\ref{sec:var-ss} were well crafted. The good agreement is also partly due to our normalization of the dark component,
where we have used the longitudinal extent of the condensate $L=\ell_1$, obtained
from the stationary solution of the full set of GPEs. Please note that the
emergence of the Faraday waves does not impact the bulk of the condensate,
and that its longitudinal extent is constant, which fully justifies our
assumption in the variational model.

\begin{figure}[!t]
\begin{center}
\includegraphics[width=6.8cm]{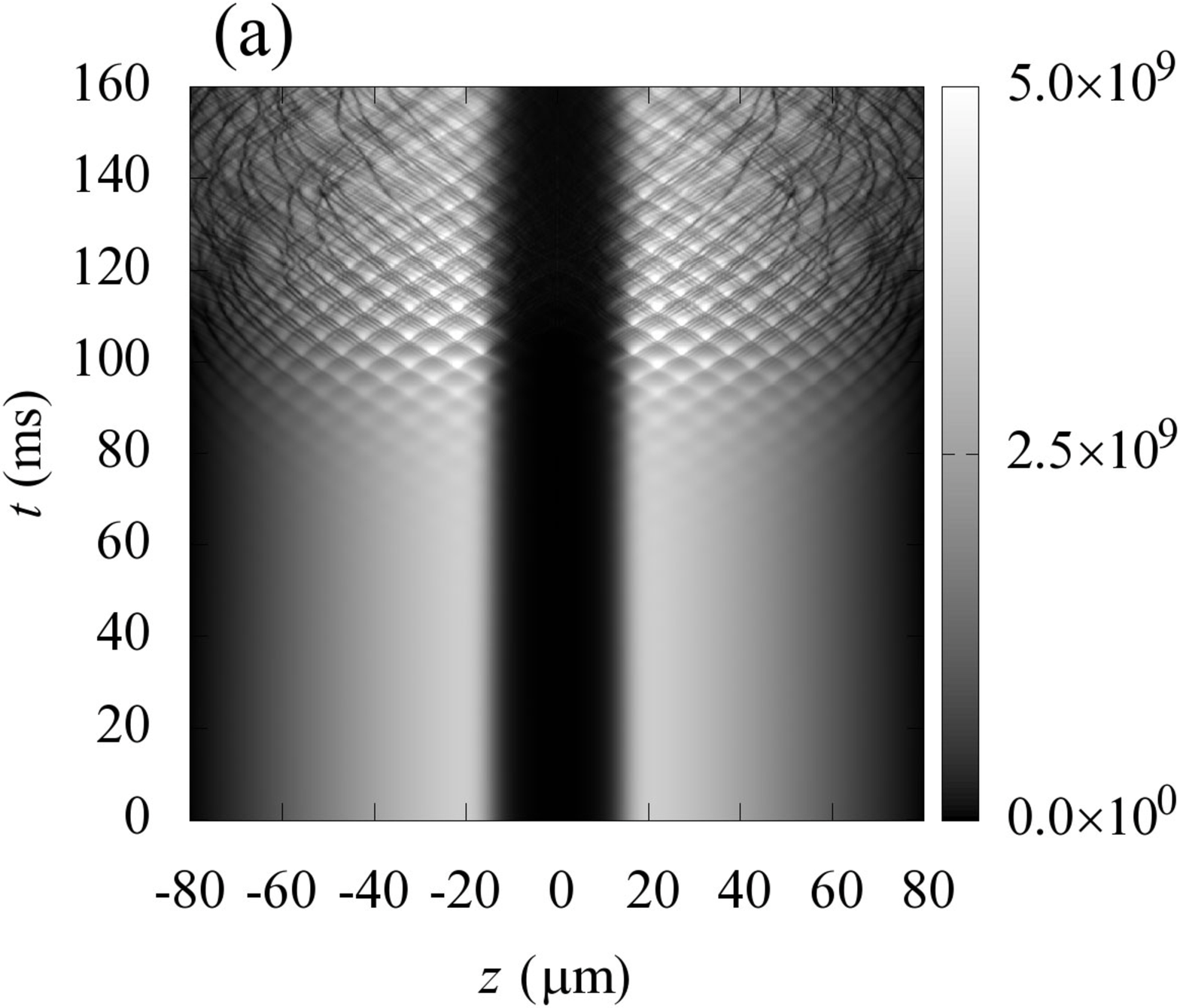}
\includegraphics[width=6.8cm]{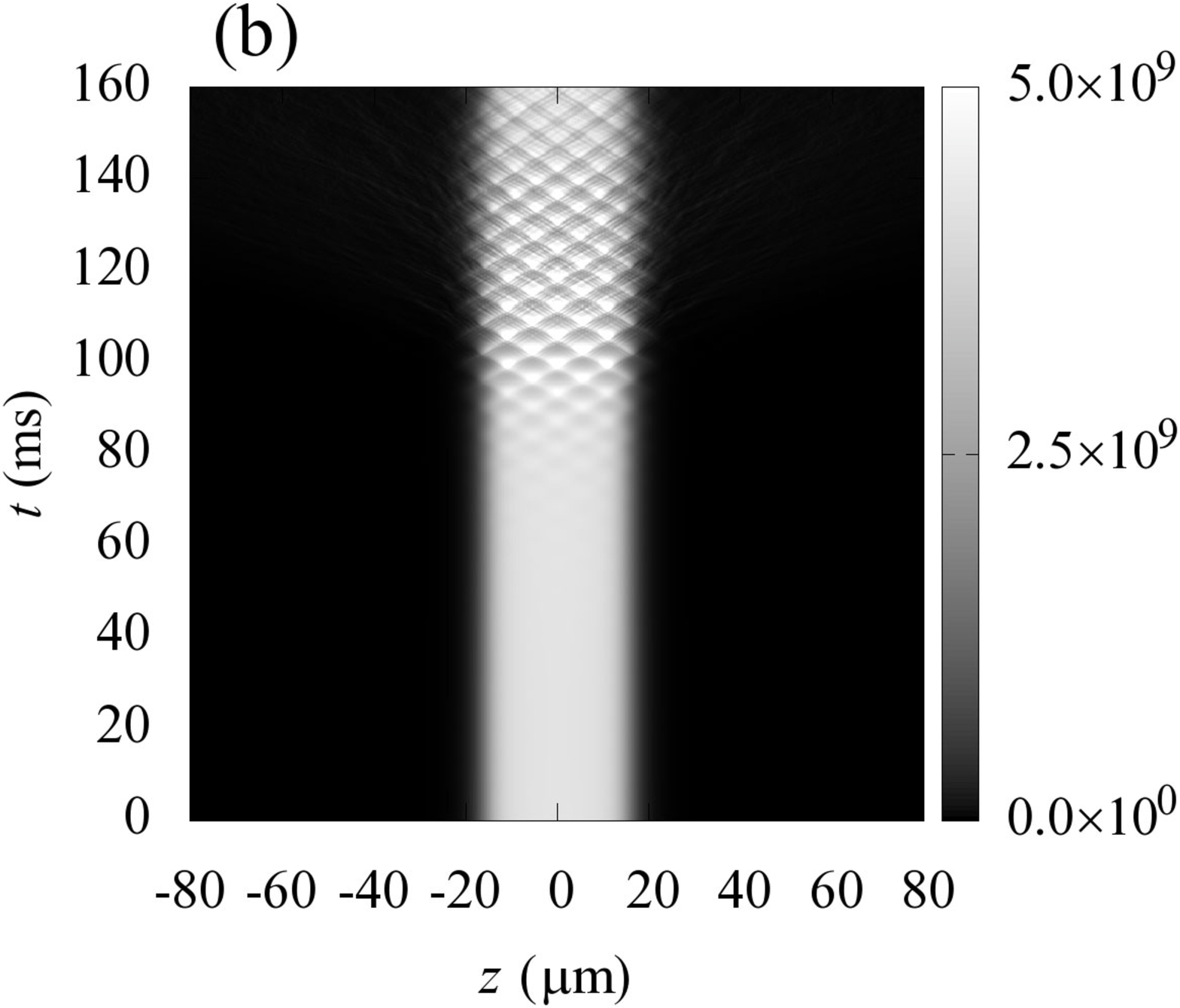}
\caption{Emergence of resonant waves in a two-component BEC system in the real-time evolution of longitudinal density profiles: (a) $n_1(z, t)$ and (b) $n_2(z, t)$. The system is initially in the symbiotic pair ground state, and is modulated with the self-resonant frequency $\omega=\Omega_{\rho 0}=160\times 2\pi$~Hz; other parameters as in Fig.~\ref{fig4}.}
\label{fig6}
\end{center}
\end{figure}

\begin{figure*}[!t]
\begin{center}
\includegraphics[width=6.5cm]{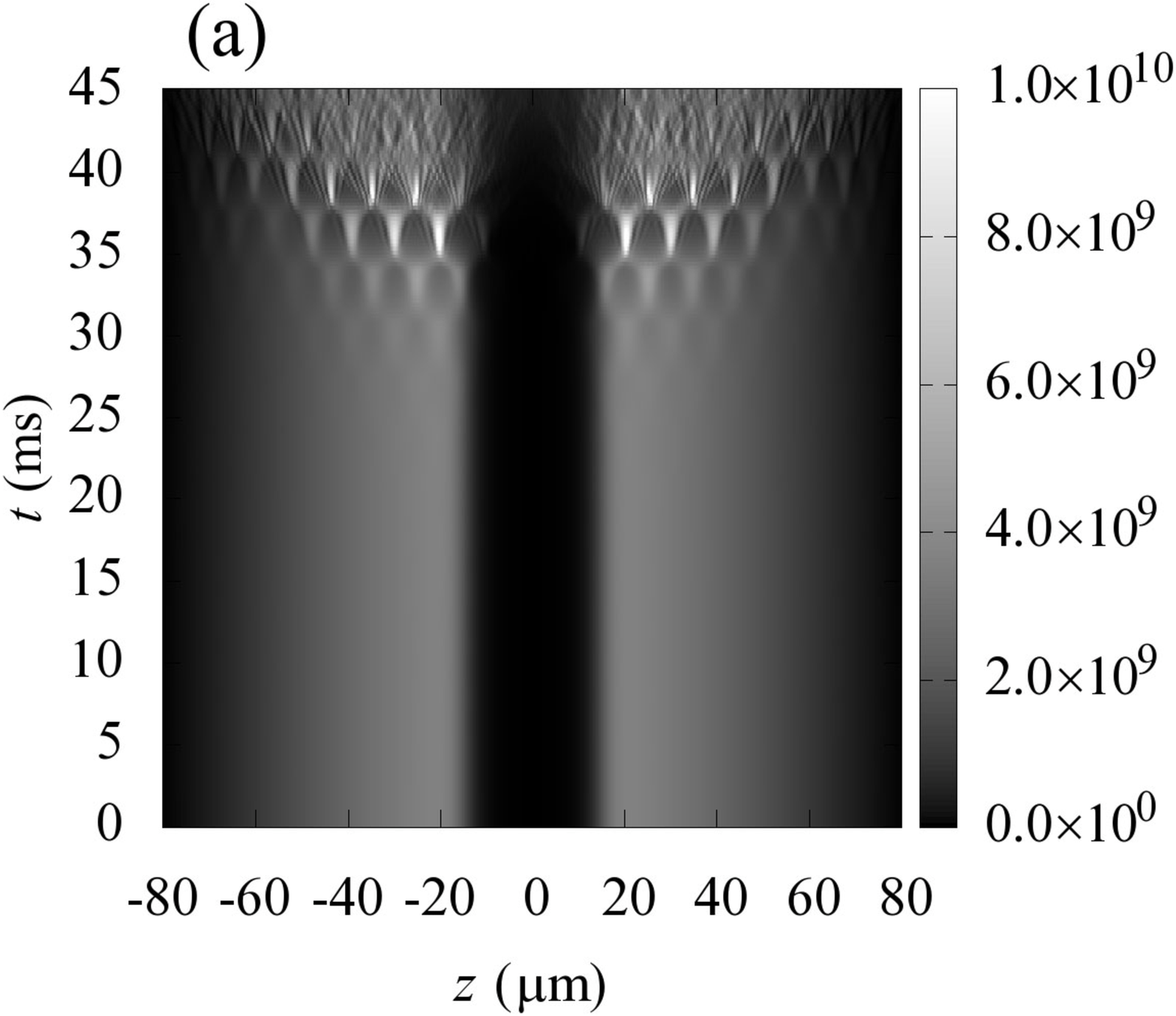}\hspace*{8mm}
\includegraphics[width=6.5cm]{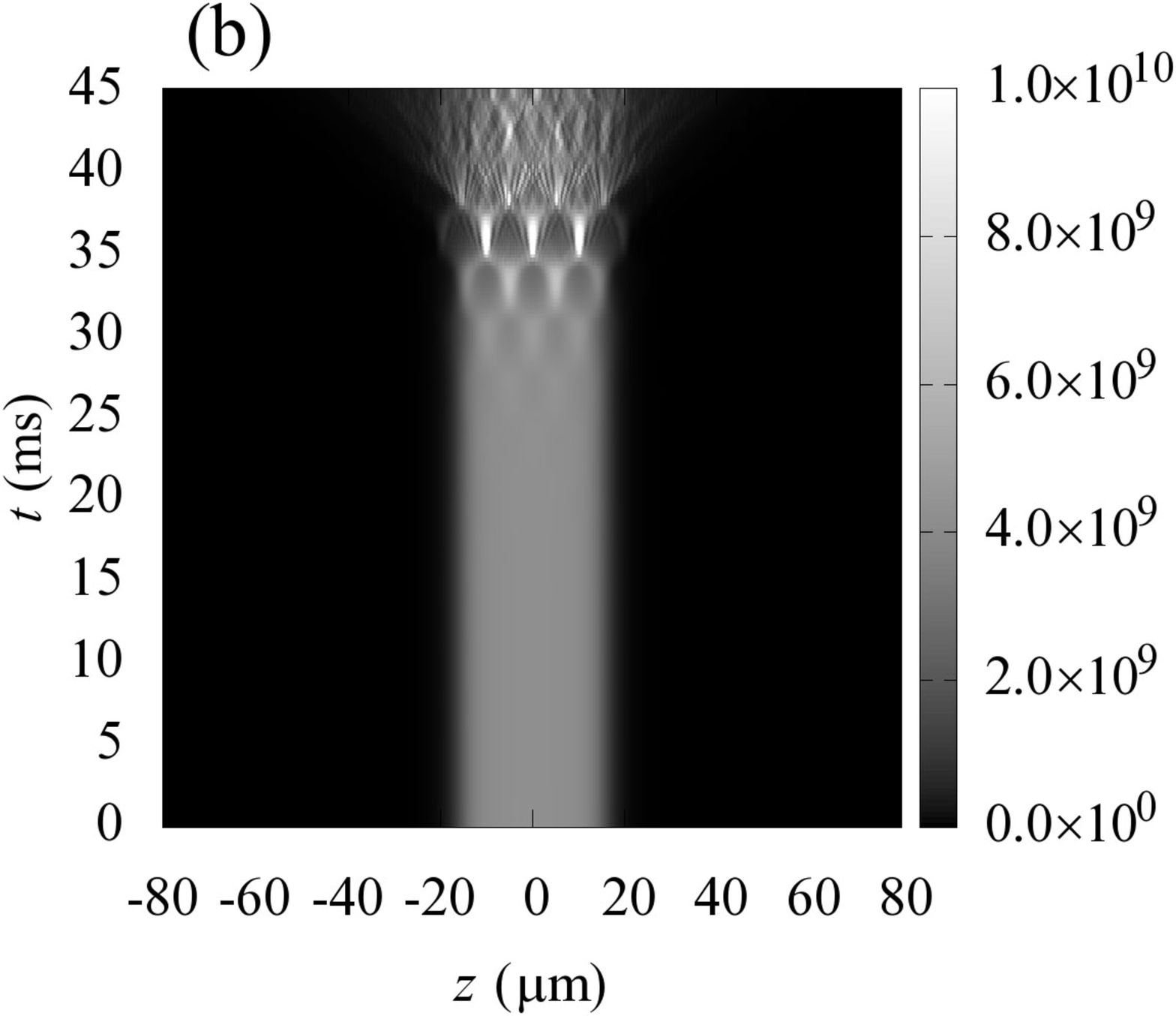}
\includegraphics[width=6.5cm]{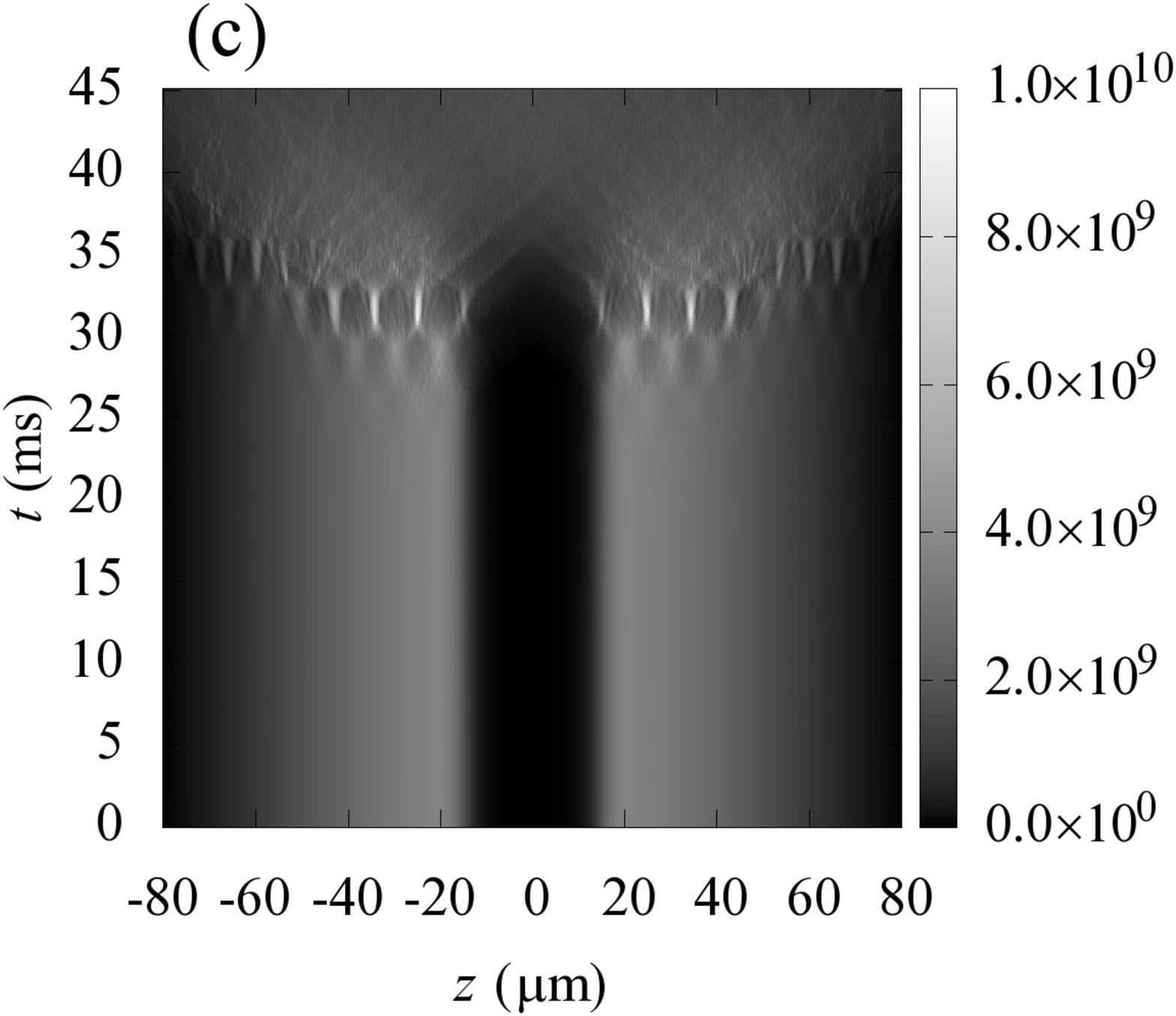}\hspace*{8mm}
\includegraphics[width=6.5cm]{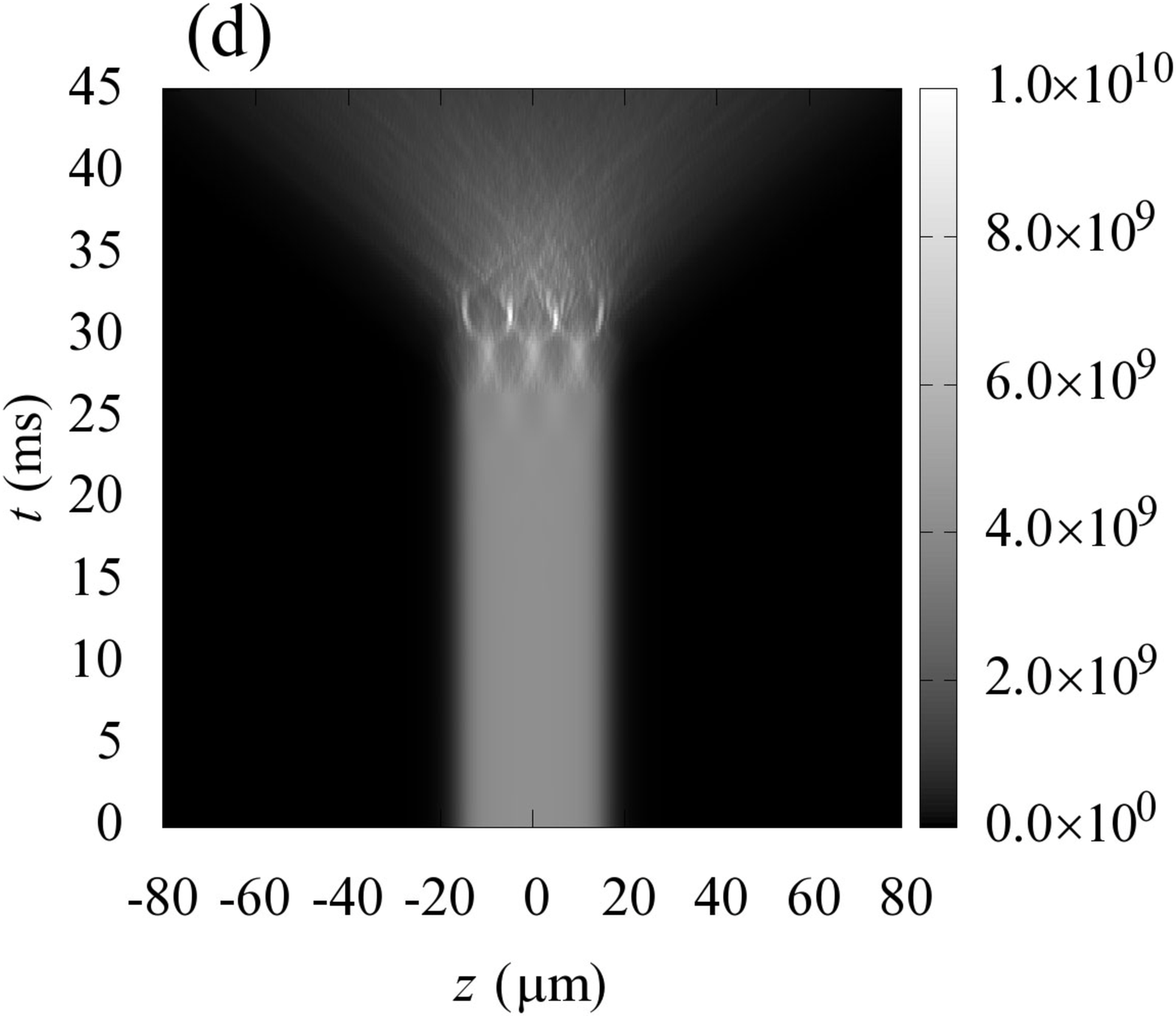}
\includegraphics[width=6.5cm]{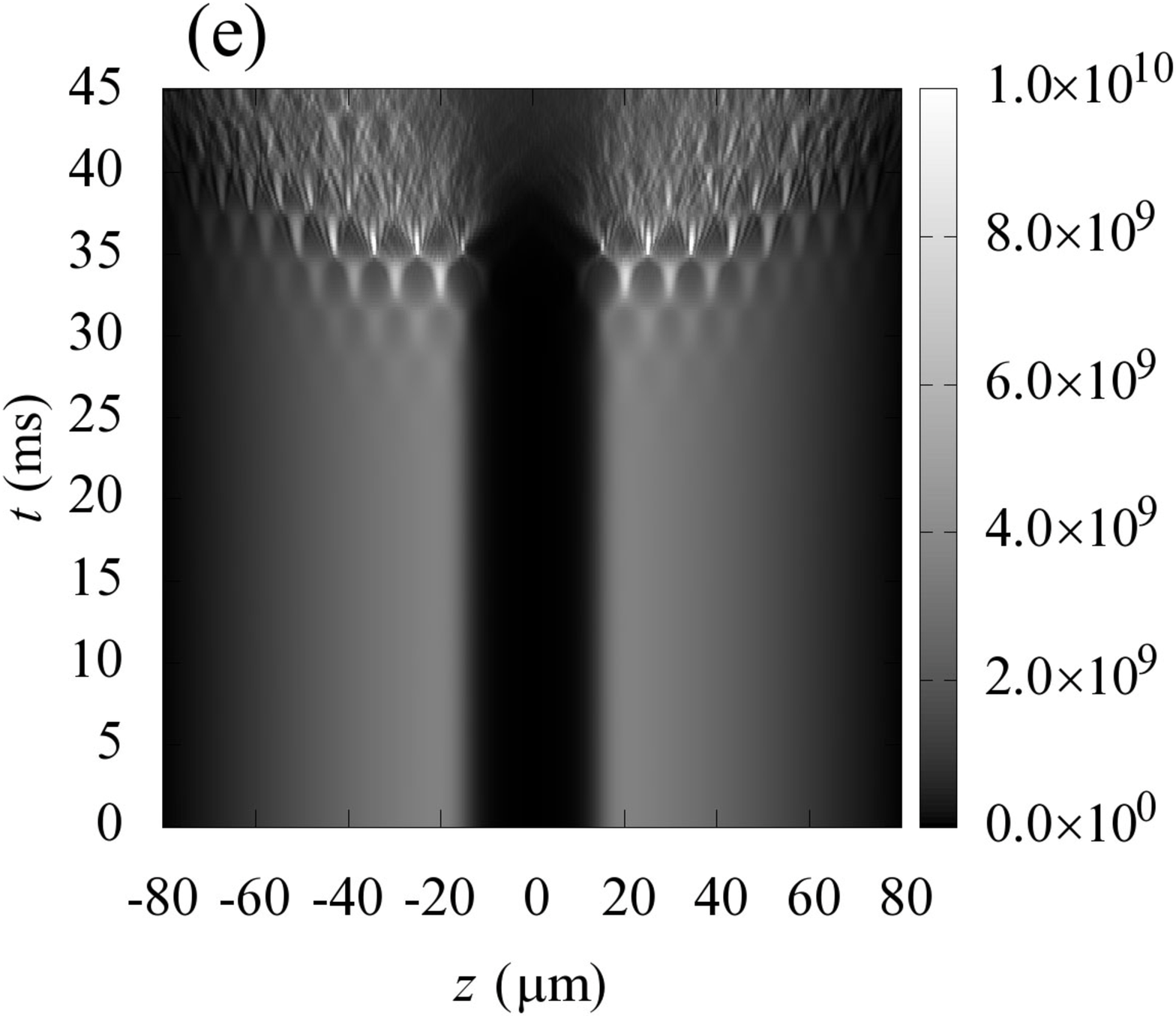}\hspace*{8mm}
\includegraphics[width=6.5cm]{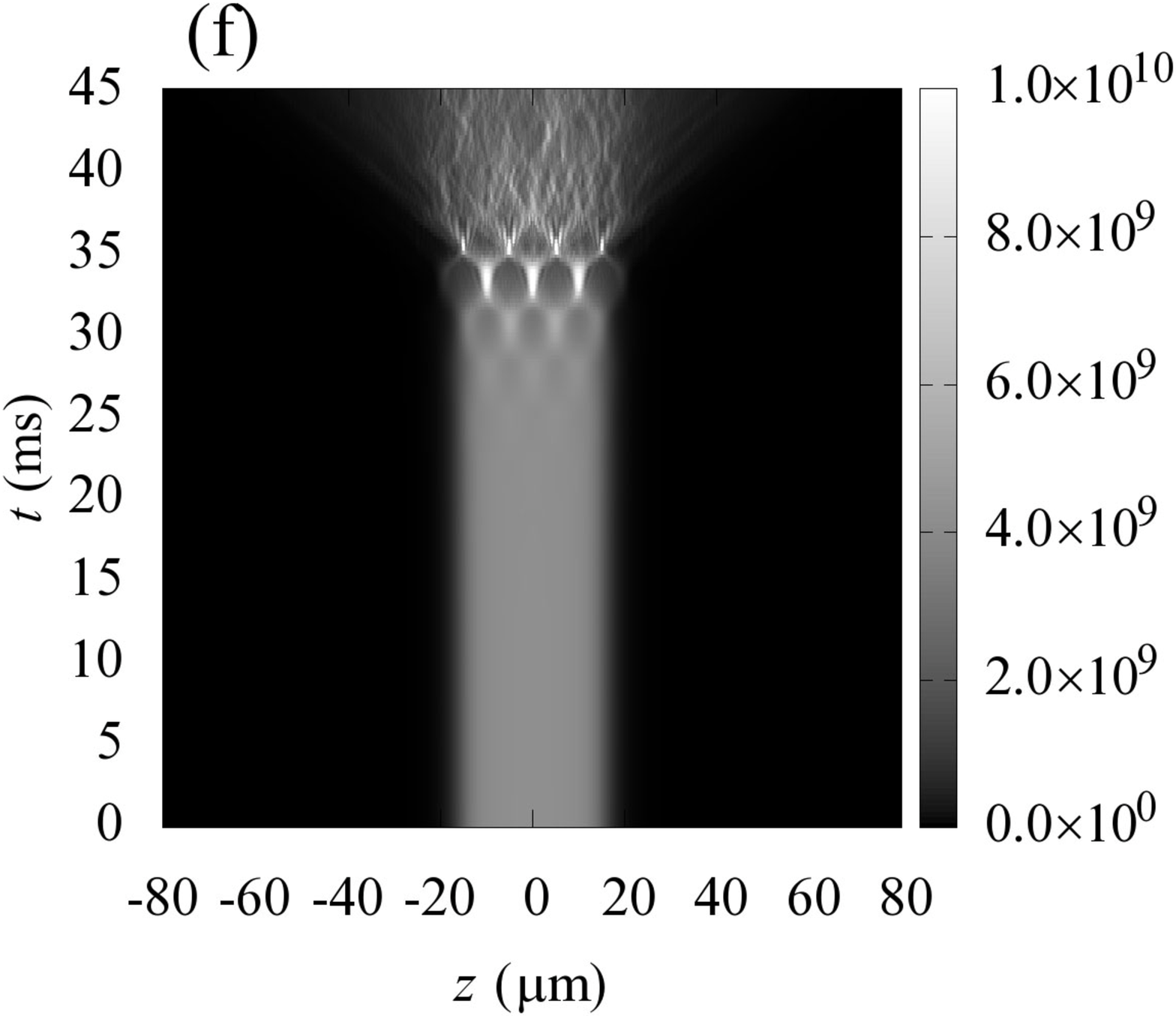}
\caption{Emergence of resonant waves in a two-component BEC system in the real-time evolution of longitudinal density profiles: $n_1(z, t)$ (left column) and $n_2(z, t)$ (right column). The system is initially in the symbiotic pair ground state, and is modulated with the frequencies: $\omega=300\times 2\pi$~Hz in (a) and (b); $\omega=2\Omega_{\rho 0}=320\times 2\pi$~Hz in (c) and (d); $\omega=340\times 2\pi$~Hz in (e) and (f). Other parameters are the same as in Fig.~\ref{fig4}.}
\label{fig7}
\end{center}
\end{figure*}

In Fig.~\ref{fig6} we show the resonant dynamics of the condensate
for a self-resonance, $\omega=\Omega_{\rho 0}=160\times 2\pi$ Hz. In this case
our numerical simulations indicate a narrow resonance where the excited
surface waves do not impact the dynamics of the bulk of the condensate. The resonant waves develop faster than the Faraday waves, and already after 70~ms are clearly visible.
However, in Fig.~\ref{fig7} we see that for a second resonance at $\omega\approx 2\Omega_{\rho 0}$, the excited surface
waves appear even much earlier, after only 25~ms, but are quite short-lived, and the collective dynamics of the two
components then takes over. This figure also illustrates that the second resonance is very broad, and covers the interval wider than $40\times 2\pi$~Hz, centered at $2\Omega_{\rho 0}=320\times 2\pi$~Hz.

\begin{figure*}[!t]
\begin{center}
\includegraphics[width=6.5cm]{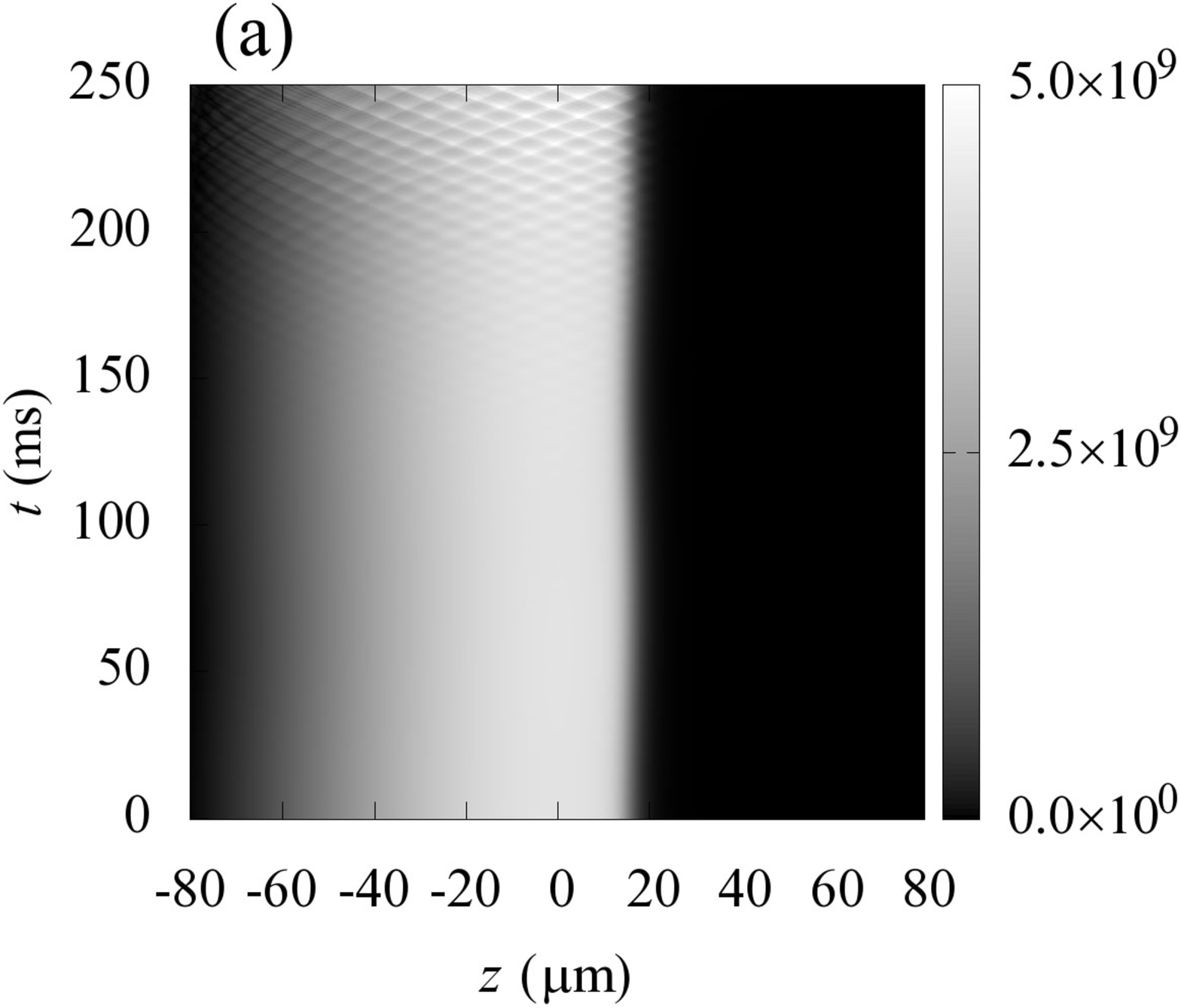}\hspace*{8mm}
\includegraphics[width=6.5cm]{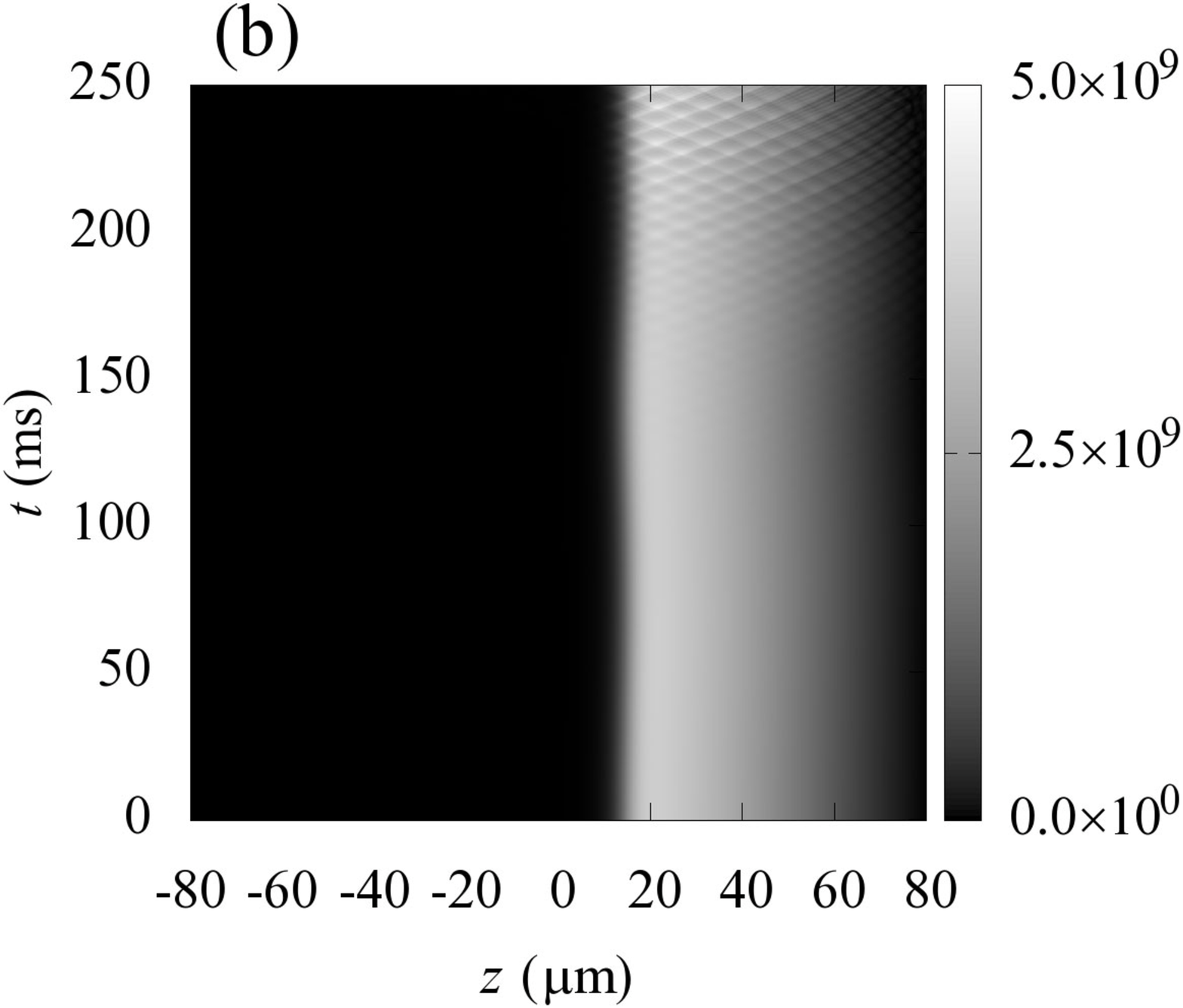}
\includegraphics[width=6.5cm]{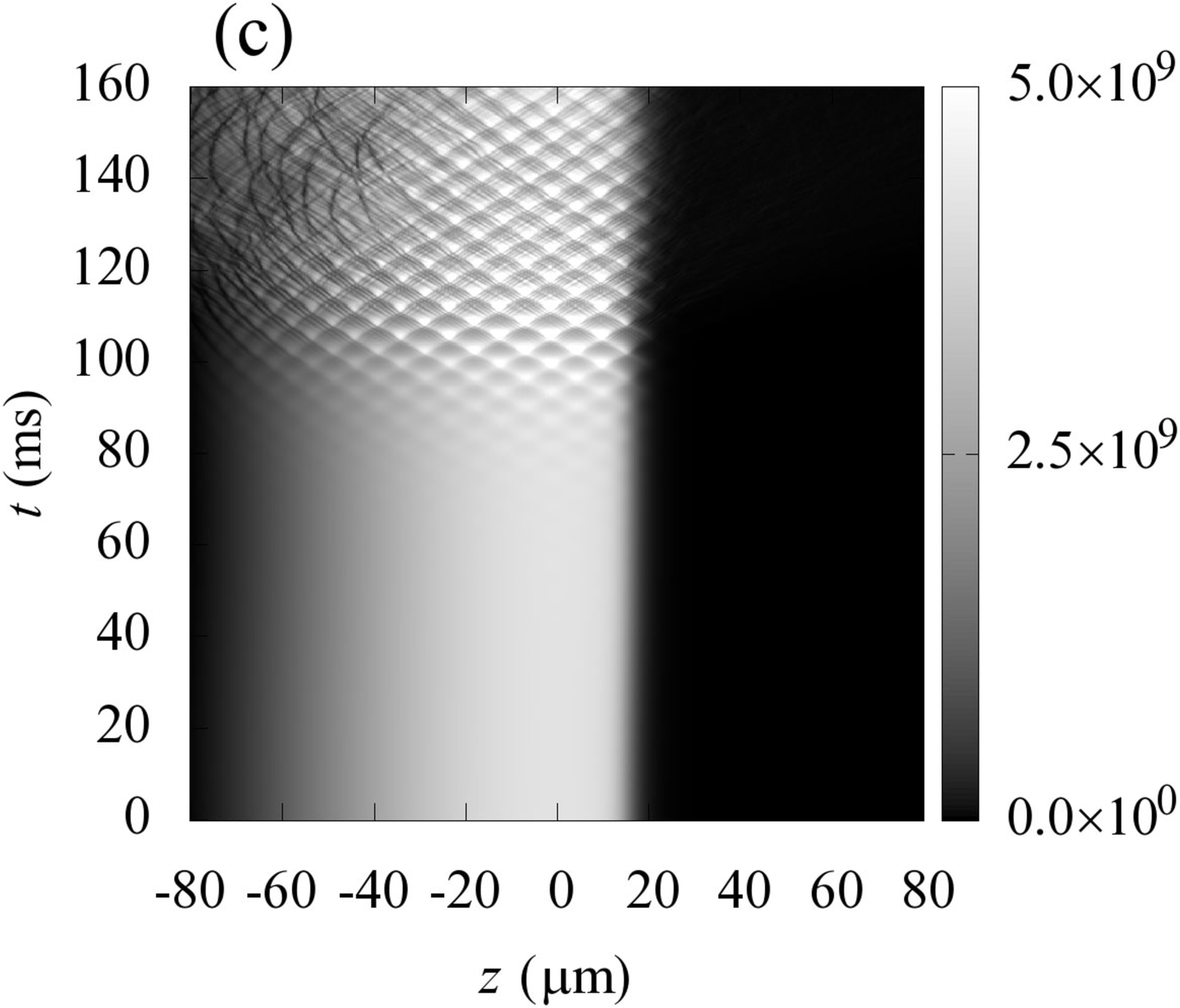}\hspace*{8mm}
\includegraphics[width=6.5cm]{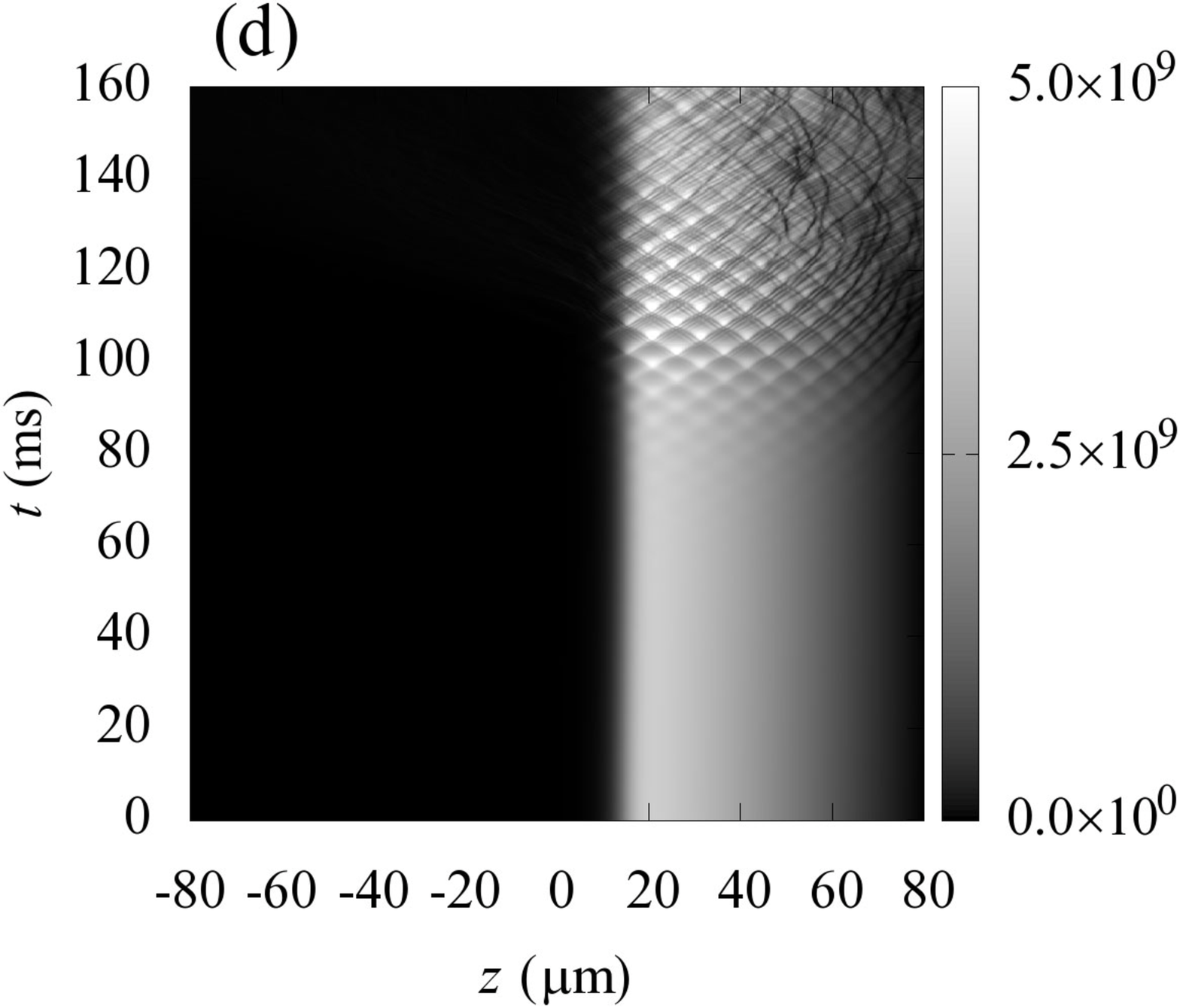}
\caption{Emergence of Faraday (a, b) and resonant (c, d) waves in a two-component BEC system in the real-time evolution of longitudinal density profiles: $n_1(z, t)$ (left column) and $n_2(z, t)$ (right column). The system is initially in the segregated excited state, and is modulated with the amplitude $\epsilon=0.1$ and frequencies: $\omega=250\times 2\pi$~Hz in (a) and (b); $\omega=\Omega_{\rho 0}=160\times 2\pi$~Hz in (c) and (d). The system contains $N_{1}=2.5\cdot 10^5$ atoms in the state A and $N_{2}=1.25\cdot 10^5$ atoms in the state B, and is confined by the trap with $\Omega_{\rho 0}=160\times 2\pi$~Hz and  $\Omega_z=7\times 2\pi$~Hz.}
\label{fig8}
\end{center}
\end{figure*}

\begin{figure}[!b]
\begin{center}
\includegraphics[width=7cm]{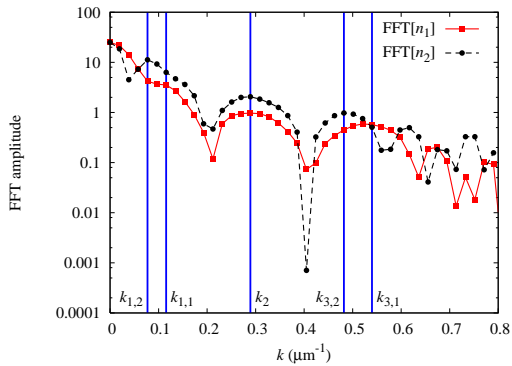}
\caption{(Color online) Fast Fourier transform of density profiles from Fig.~\ref{fig8} for two condensate components at $t=200$~ms.}
\label{fig9}
\end{center}
\end{figure}

We note, in particular, that precisely at the
second resonance the two components explode into one another (due
to the resonant energy transfer), thereby becoming effectively miscible.
This resonant transition to miscibility is specific to two-component
systems, and has also been recently reported in binary dipolar BECs
\cite{TransToMisc}. Our analytical treatment of the surface waves
indicates a period of $9.3\,\mu$m for the self-resonance, while the full GPEs numerically yield periods of $9.3\, \mu$m for the first component and $9.0\, \mu$m for the second component, which is again an excellent agreement. For the broad resonance at $\omega\approx 2\Omega_{\rho 0}$, the variational treatment gives $4.9\, \mu$m, while numerically we find periods of $9.6\, \mu$m for both components. Here the agreement is only qualitative, due to a violent dynamics observed numerically, which cannot be fully captured by simple ans\"atze used in Sec.~\ref{sec:var-ss}.

As symbiotic states are routinely obtained in $^{87}$Rb condensates
\cite{DBSolitonFirstExp,DarkBrightPhysLettA}, the observation of the
Faraday waves is definitely within the current experimental capabilities,
as is the observation of self-resonant waves at $\omega=\Omega_{\rho 0}$.
The observation of resonant waves for $\omega$ close to the second resonance $2\Omega_{\rho 0}$
seems, however, unlikely, as these waves quickly fade out in favor
of a forceful resonant dynamics that takes the condensate into the
miscible regime and can, for longer timescales, turn the condensate
unstable.

\subsection{Segregated states}
\label{sec:res-seg}

\begin{figure*}[!t]
\begin{center}
\includegraphics[width=6.6cm]{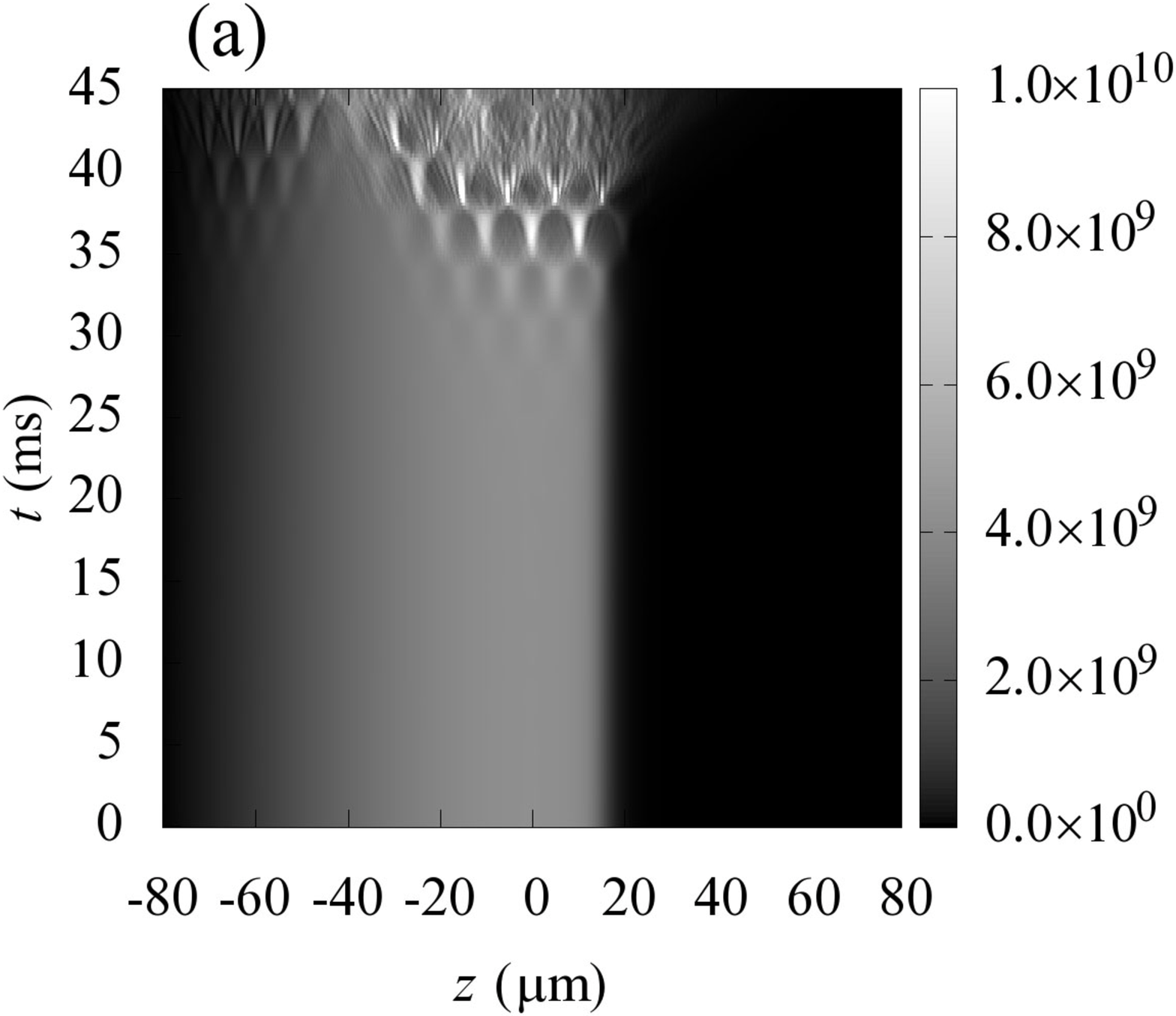}\hspace*{8mm}
\includegraphics[width=6.6cm]{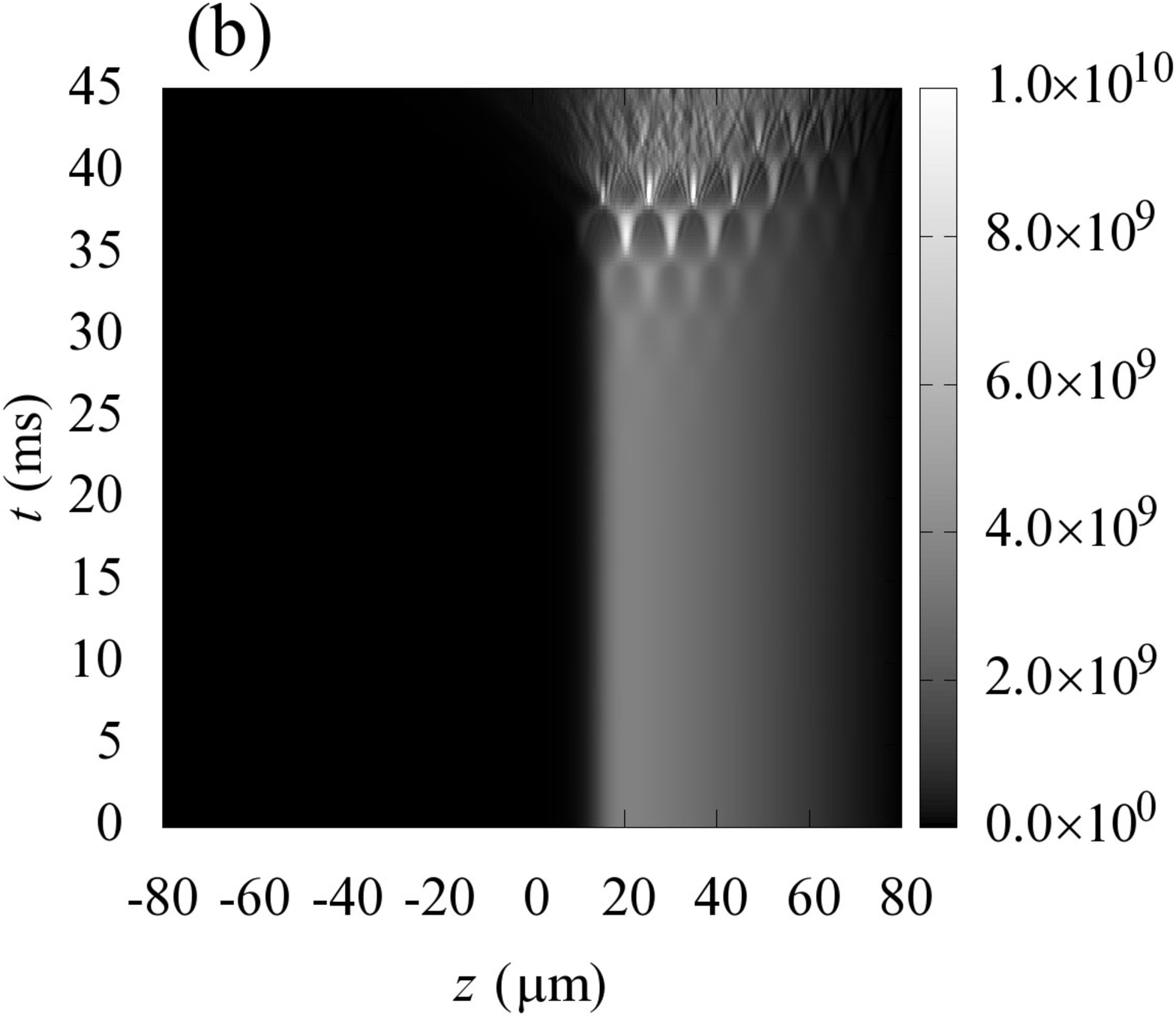}
\includegraphics[width=6.6cm]{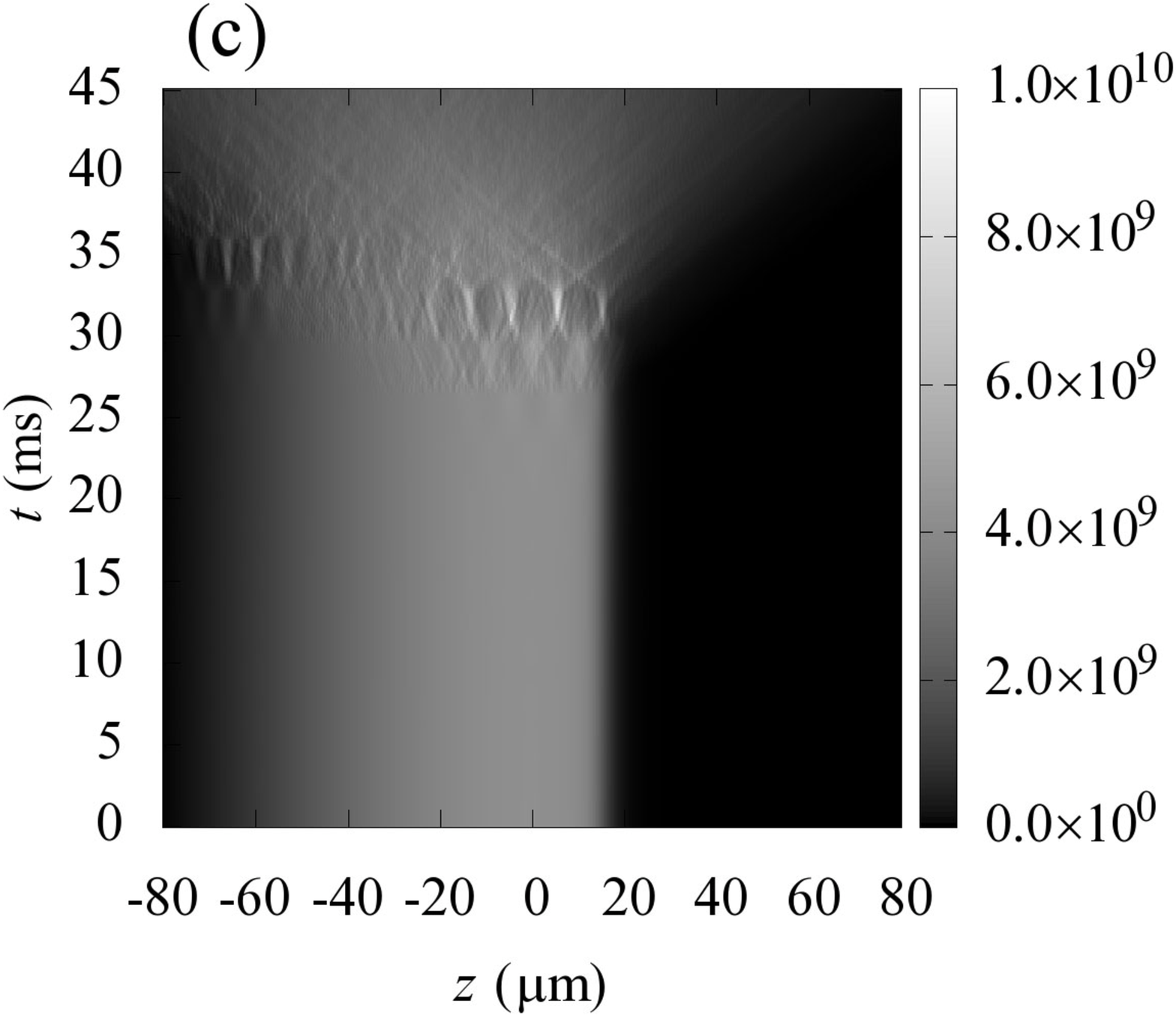}\hspace*{8mm}
\includegraphics[width=6.6cm]{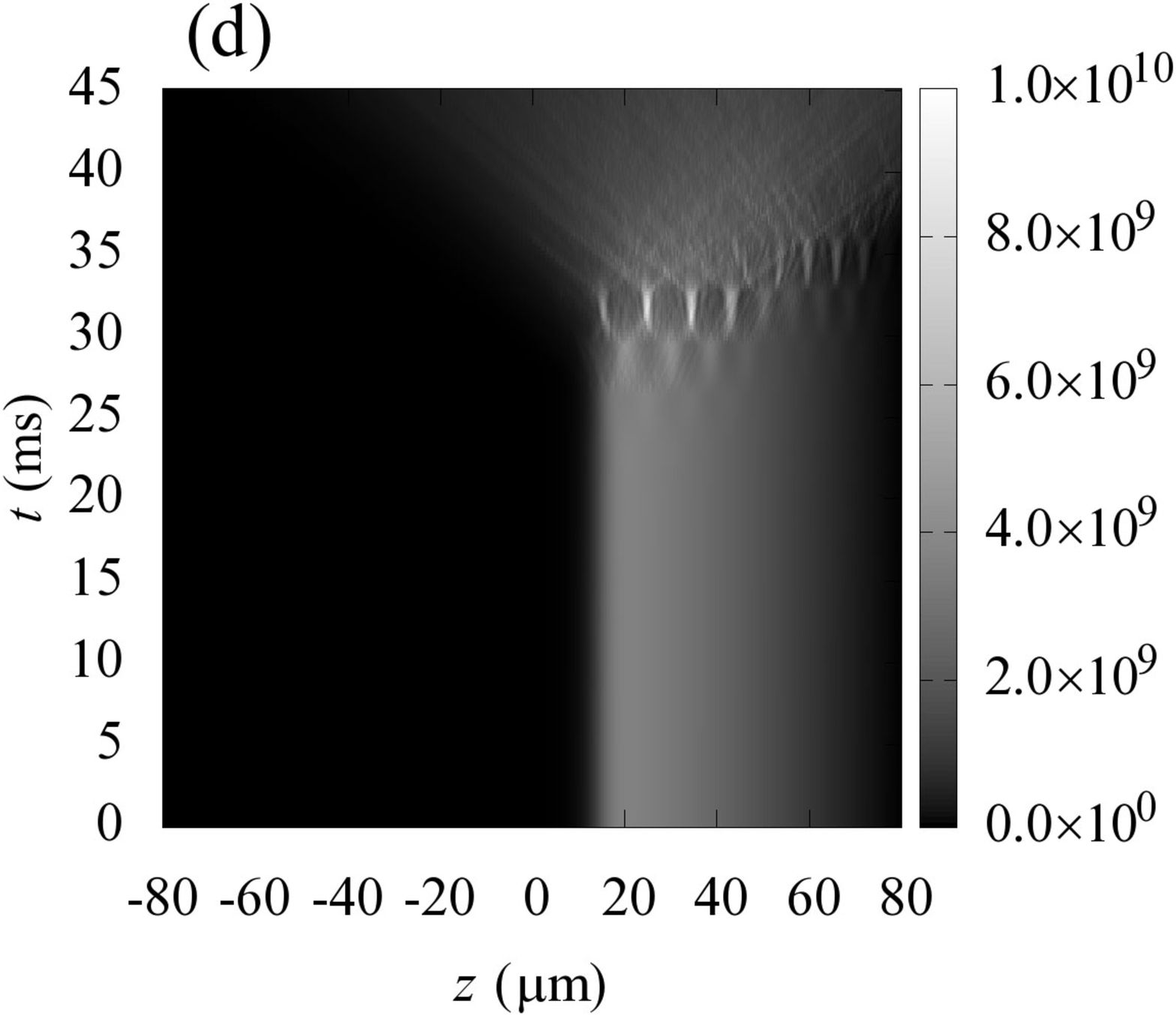}
\includegraphics[width=6.6cm]{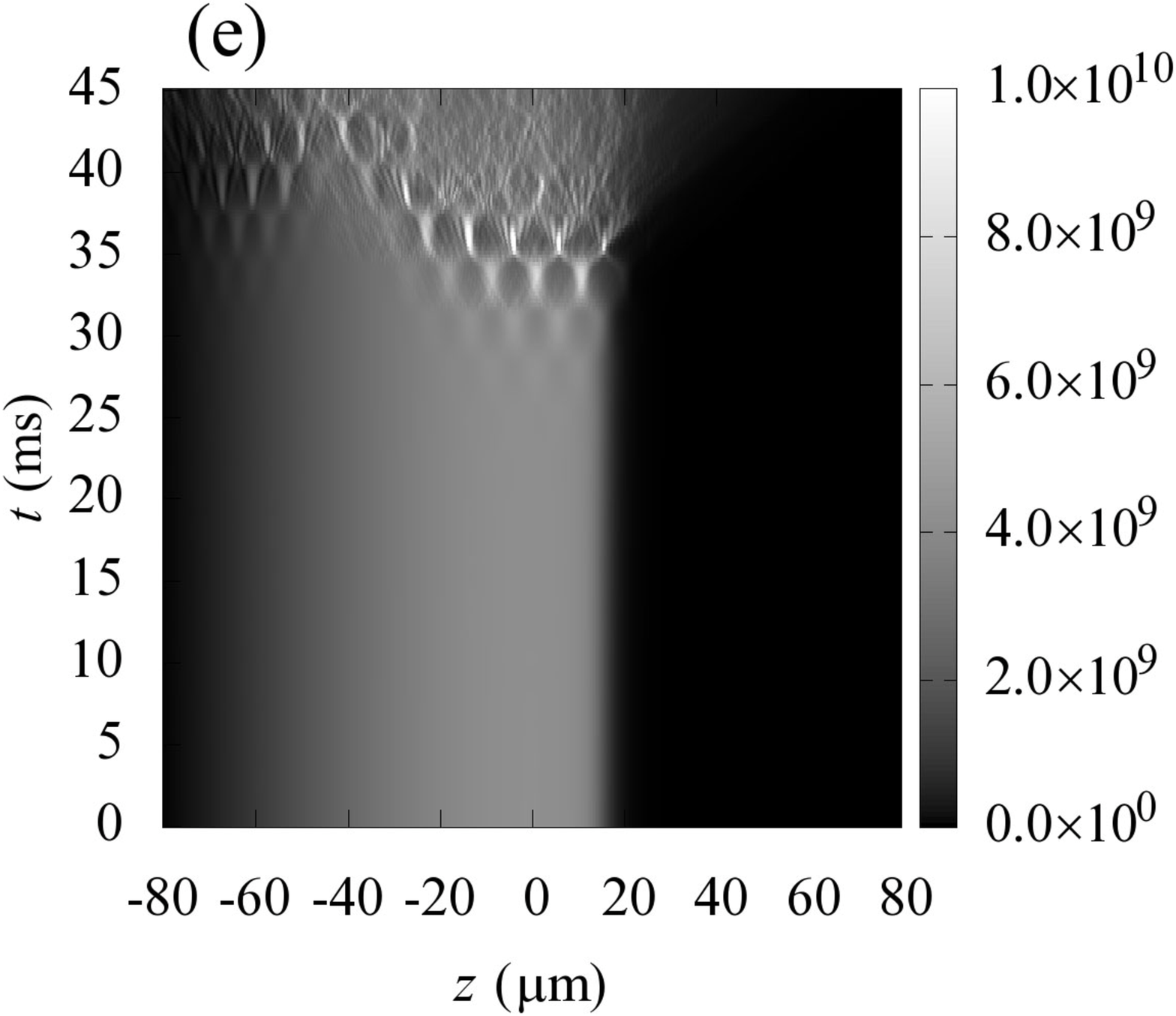}\hspace*{8mm}
\includegraphics[width=6.6cm]{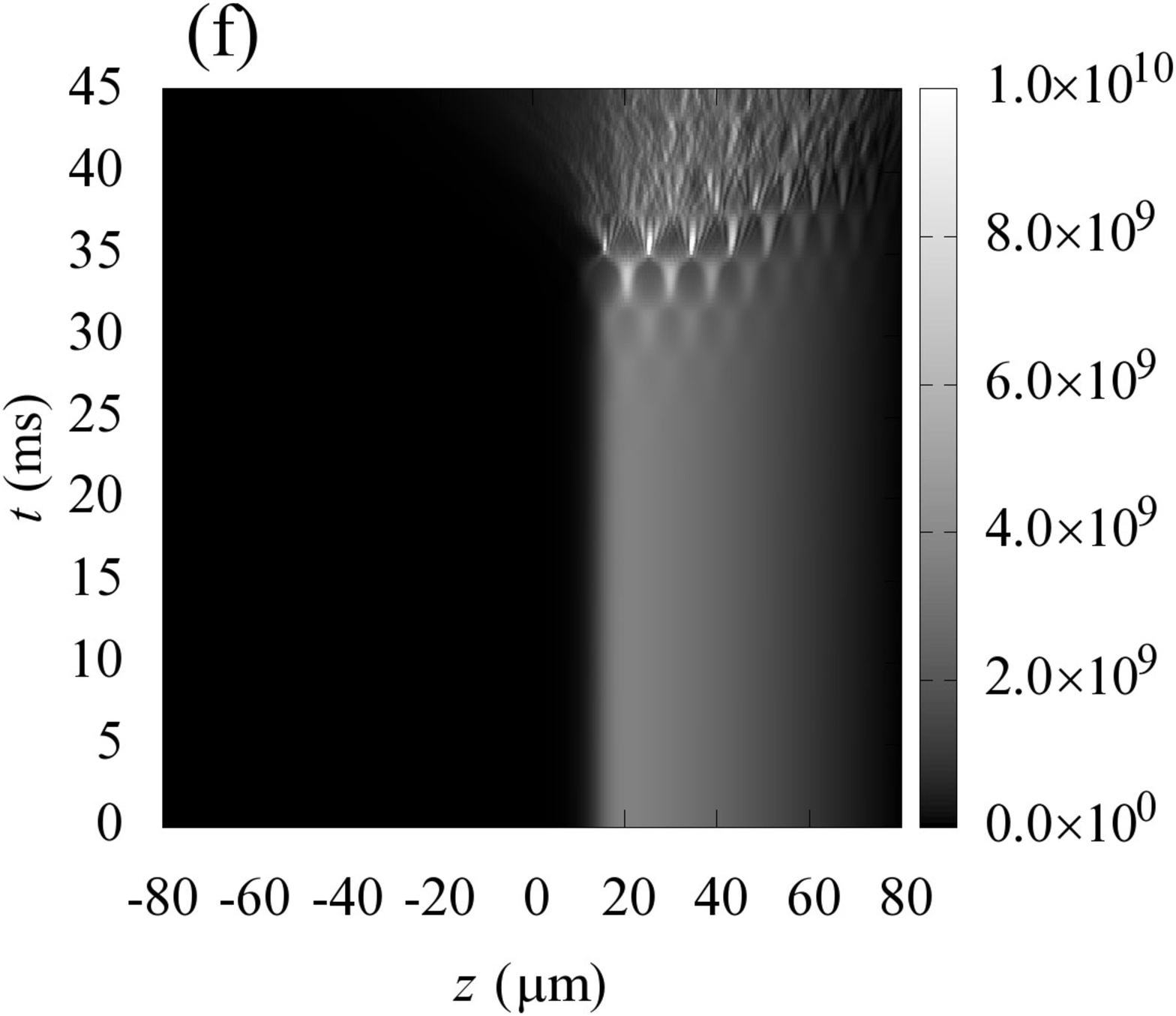}
\caption{Emergence of resonant waves in a two-component BEC system in the real-time evolution of longitudinal density profiles: $n_1(z, t)$ (left column) and $n_2(z, t)$ (right column). The system is initially in the segregated excited state, and is modulated with the frequencies: $\omega=300\times 2\pi$~Hz in (a) and (b); $\omega=2\Omega_{\rho 0}=320\times 2\pi$~Hz in (c) and (d); $\omega=340\times 2\pi$~Hz in (e) and (f). Other parameters are the same as in Fig.~\ref{fig8}.}
\label{fig10}
\end{center}
\end{figure*}

In Figs.~\ref{fig8}a and \ref{fig8}b and we show the formation of Faraday waves at $\omega=250\times 2\pi$~Hz for a segregated initial state using the same experimental setup as in the previous case. The waves again appear after roughly 150~ms and we show
in Fig.~\ref{fig9} the Fourier spectrum of the longitudinal density profile at $t=200$~ms. The spectrum's two peaks $k_{1,1}$ and $k_{1,2}$ correspond to longitudinal extents of components, $81.5\, \mu$m and $54.3\,\mu$m, while the peak $k_2$ corresponds to the width of the interface area between the components of $21.7\, \mu$m.

Periods of the Faraday waves are determined by the peaks $k_{3,1}$ and $k_{3,2}$, and have the values $11.6\, \mu$m for the first and $13.0\, \mu$m for the second component. The dispersion relation derived in Sec.~\ref{sec:var-seg} indicates a period of $16.8\, \mu$m, which overestimates the previous result by roughly 30\% due to several reasons: the ans\"atze we use in  Sec.~\ref{sec:var-seg} for two components are symmetric around $z=0$, they have the same normalization, and also they exhibit the Gaussian exponential cut-off away from the center of the trap.
From Fig.~\ref{fig3} we already see that these are simplifications made only to ensure analytical tractability of the variational calculation. 
We also stress that both components are in the Thomas-Fermi regime, hence their wave functions are well-localized and show a distinct quasi-singular cut-off at borders of the cloud, not the Gaussian-type tail. However, even with these simplifications, the variational approach is able to give reasonable estimate for the period of Faraday waves.

We also numerically study resonant dynamics of the two-component BEC system for the self-resonance at $\omega=\Omega_{\rho 0}$. In Figs.~\ref{fig8}c and \ref{fig8}d we show real-time evolution of density profiles for both components. The resonance is relatively narrow and the excited surface wave do
not impact the overall dynamics of the condensate. The resonant surface waves appear after around 80~ms, and have the period of $11.2\, \mu$m in the first and $12.5\, \mu$m in the second component, while the variational approach, Eq.~(\ref{eq:kFseg}), gives the period of $13.1\, \mu$m. The agreement is quite good, but only coincidental, since we know that the ans\"atze used in Sec.~\ref{sec:var-seg} are oversimplified, and that we cannot expect more than an order of magnitude agreement.

In the case of a second resonance at $\omega=2\Omega_{\rho 0}$, the excited surface waves fade out in favor of the forceful resonant
dynamics that turns the two components miscible. The resonant waves appear already after around 25~ms, much faster than for the self-resonance. As in the case of the symbiotic pair state, the second resonance is much broader than the self-resonance, with the width of more than $40\times 2\pi$~Hz, as illustrated in Fig.~\ref{fig10}. Our analytical treatment of the surface waves indicates a period of $6.6\, \mu$m, while the Fourier analysis of the full numerical solution of GPEs yield periods of $10.2\, \mu$m in the first and $9.6\, \mu$m in the second component. As expected, the variational treatment gives a fair estimate of the period of resonant surface waves.

\section{Conclusions}
\label{sec:con}

We have shown the emergence of Faraday waves in binary non-miscible
BECs subject to periodic modulations of the transverse confinement.
Considering the $\left|1,-1\right\rangle $ and $\left|2,1\right\rangle $
hyperfine states of realistic $^{87}$Rb condensates, we have shown
that there are two types of experimentally relevant stationary configurations: a symbiotic
pair ground state, where one component is trapped by the other, and a configuration where the components are
well segregated, which represents a first excited state of the system. For both types
of configurations we have shown by extensive numerical study that the surface
waves in the two components emerge simultaneously, are of similar periods
and do not impact the dynamics of the bulk of condensate far from
resonances. We have derived analytically periods of the excited surface
waves using first a variational treatment of the full system of coupled GPEs that reduced
the dynamics of the condensate to a set of coupled ordinary differential
equations, and then using a Mathieu-type analysis of these equations we have determined
the most unstable solutions. The obtained analytical results for periods of surface waves are
shown to give excellent estimates by comparison with the Fourier analysis of numerical solutions of the full set of GPEs.
The resonant dynamics seen for modulations equal to the radial trapping frequency (self-resonance), as well as to twice this value (second resonance) were investigated numerically, and we have found that the emergent resonant surface waves appear much faster than the Faraday waves. Using the variational approach developed in Sec.~\ref{sec:var}, the periods of resonant surface waves can be also estimated, and are found to be in reasonable agreement with the numerical results. In the case of a second resonance, the emergent
surface waves for both types of initial configurations are found to fade out in favor of a forceful resonant dynamics that
turns the binary condensate miscible. This is an interesting phenomenon, and we believe that GPE framework captures its dynamics quite well, similarly to e.g.~Ref.~\cite{collision}, where the condensate was split into two separated regions and then the collisional dynamics of the two halves was observed and found to be in excellent agreement with GPE predictions.

Extending the present investigation to experimental setups with anisotropic
transverse confinement (such as the one in Ref. \cite{DarkBrightPhysLettA}) presents
one of interesting venues for future studies. As fully three-dimensional
simulations are extremely time-consuming, a natural first step would
be to extend the existing (cylindrically symmetric) non-polynomial
Schr\"{o}dinger equations \cite{salanischboris,MateoDelgado} to
cover the dynamics of binary BECs subject to anisotropic transverse
confinement. Also in this direction the study of predominantly two-dimensional
(pancake-shaped) condensates has the potential to uncover a rich variety of patterns,
and one could even expect the formation of spatio-temporal chaos in such systems.
Finally, another natural extension of this work would be to consider parametric 
resonances and the ensuing Faraday waves in miscible condensates. In this setting the dynamics
of the waves in the two components can be mapped variationally onto
a set of coupled Mathieu equations whose most unstable solutions are,
however, not known analytically.

\begin{acknowledgments}
We would like to thank Peter Engels for many insightful suggestions and careful reading of the manuscript. We also thank Panos Kevrekidis, Dumitru Mihalache, Mihnea Dulea, Madalin Bunoiu, and Ivana Vidanovi\' c for useful discussions. A.I.N.~acknowledges support from CNCS-UEFISCDI through the postdoctoral grant PD122 contract No.~35/28.07.2010 and project PN09370104 financed by ANCS, while A.B. acknowledges support from the Serbian Ministry of Education and Science under project No.~ON171017, and from the European Commission under EU FP7 projects PRACE-1IP, PRACE-2IP, HP-SEE and EGI-InSPIRE.
\end{acknowledgments}

\end{document}